\documentclass[prb,showpacs,twocolumn,aps,superscriptaddress,floatfix]{revtex4}
\usepackage{amsmath}
\usepackage{amssymb}
\usepackage{bm}
\usepackage{graphicx}
\usepackage{color}
\usepackage{ulem}
\usepackage{hyperref}
\usepackage{comment}

\DeclareMathOperator{\Tr}{Tr}

\begin{document}

\title{
Electronic spectrum of twisted bilayer graphene
}

\author{A.O. Sboychakov}
\affiliation{CEMS, RIKEN, Wako-shi, Saitama,
351-0198, Japan}
\affiliation{Institute for Theoretical and Applied Electrodynamics, Russian Academy of Sciences, 125412 Moscow, Russia}

\author{A.L. Rakhmanov}
\affiliation{CEMS, RIKEN, Wako-shi, Saitama,
351-0198, Japan}
\affiliation{Institute for Theoretical and Applied Electrodynamics, Russian Academy of Sciences, 125412 Moscow, Russia}
\affiliation{Moscow Institute of Physics and Technology, Dolgoprudny, Moscow Region, 141700 Russia}
\affiliation{All-Russia Research Institute of Automatics, Moscow, 127055 Russia }

\author{A.V. Rozhkov}
\affiliation{CEMS, RIKEN, Wako-shi, Saitama,
351-0198, Japan}
\affiliation{Institute for Theoretical and Applied Electrodynamics, Russian Academy of Sciences, 125412 Moscow, Russia}
\affiliation{Moscow Institute of Physics and Technology, Dolgoprudny, Moscow Region, 141700 Russia}

\author{Franco Nori}
\affiliation{CEMS, RIKEN, Wako-shi, Saitama,
351-0198, Japan}
\affiliation{Department of Physics, University of Michigan, Ann
Arbor, MI 48109-1040, USA}

\begin{abstract}
We study the electronic properties of twisted bilayers graphene in the
tight-binding approximation. The interlayer hopping amplitude is  modeled
by a function, which depends not only on the distance between two carbon
atoms, but also on the positions of neighboring atoms as well. Using the
Lanczos algorithm for the numerical evaluation of eigenvalues of large
sparse matrices, we calculate the bilayer single-electron spectrum for
commensurate twist angles in the range
$1^{\circ}\lesssim\theta\lesssim30^{\circ}$.
We show that at certain angles $\theta$ greater than
$\theta_{c}\approx1.89^{\circ}$
the electronic spectrum acquires a finite gap, whose value could be as large
as
$80$\,meV.
However, in an infinitely large and perfectly clean sample the gap as a
function of $\theta$ behaves non-monotonously, demonstrating exponentially-large jumps for very small variations of $\theta$. This sensitivity to
the angle makes it impossible to predict the gap value for a given
sample, since in experiment $\theta$ is always known with certain error. To
establish the connection with experiments, we demonstrate that for a
system of finite size
$\tilde L$
the gap becomes a smooth function of the twist angle. If the sample is
infinite, but disorder is present, we expect that the electron
mean-free path plays the same role as
$\tilde L$.
In the regime of small angles
$\theta<\theta_c$,
the system is a metal with a well-defined Fermi surface which is reduced to
Fermi points for some values of $\theta$. The density of states in the
metallic phase varies smoothly with $\theta$.
\end{abstract}


%
%
%
%
%
%
%
%
%
%

\maketitle

\section{Introduction}

Bilayer graphene is attracting considerable
attention. Recent experimental studies (including scanning
tunneling microscopy~\cite{STM_LL,STMNanoLett,STM_VHS,STM_VHS2},
Raman~\cite{Raman1,Raman2}, and angular resolved photoemission
spectroscopy~\cite{ARPES_PRB,ARPES_NatMat})
revealed that, in many cases, the structure of bilayer samples is far from
the ideal AB stacking and is
characterized by a non-zero twist angle $\theta$ between graphene layers.
The physics of twisted bilayer graphene (tBLG) is very rich. The system
demonstrates Dirac spectrum with a $\theta$-dependent Fermi
velocity~\cite{STM_LL,Raman1},
low-energy van Hove singularities~\cite{STM_VHS,STM_VHS2},
and other interesting
features~\cite{Hall1,Hall2}.

The theoretical description of the low-energy electronic
properties of twisted bilayer graphene is based on the notion that for certain, so-called
`commensurate', values of $\theta$, the tBLG lattice may be thought of as a
periodic repetition of supercells, containing large number of carbon atoms.
For such angles, numerical studies based on density functional theory and
tight-binding
calculations~\cite{Pankratov1,Pankratov2,Pankratov3,Pankratov4,NanoLett,
TramblyTB_Loc,Morell}
were performed. Since the number of atoms in an elementary unit cell of the
tBLG superlattice may be quite substantial, especially at small twist
angles, the
{\it ab initio}
calculations incur a significant computational cost. Therefore, the use
of such approaches is quite limited. To avoid this difficulty, several
semi-analytical theories have been developed for describing the low-energy
electronic properties of the
tBLG~\cite{MeleReview,dSPRL,dSPRB,PNAS,MelePRB2010,MelePRB2011,
NonAbelianGaugePot}.

These low-energy theories operate mainly with the electronic states near
the Dirac cones, which the tBLG inherits from its two constituent layers.
In the tBLG, the Dirac cones with equal chirality are
located close to each other in momentum space.
The interlayer hopping couples these cones and suppresses
the Fermi velocity~\cite{STM_LL,Raman1},
which becomes a function of $\theta$.

If subtler effects are of
interest~\cite{MelePRB2010},
a term hybridizing these Dirac cones must be added to the effective
long-wave Hamiltonian of the tBLG. The corresponding electronic spectrum
obtained is gapped or gapless depending on the type of commensurate
structure.

When the twist angle is small
($\theta\lesssim2^{\circ}$),
the electronic structure changes qualitatively. The picture with Dirac
cones becomes irrelevant. Instead, the system acquires a finite density of
states at the Fermi
level~\cite{dSPRB}.

Yet, despite definite progress, several important theoretical issues remain
unaddressed. For example, the regime of low-twist angles received very limited
attention. The regime of larger angles was studied in more details. However,
the current understanding of this limit is not without discrepancies. The types of
spectra predicted in
Ref.~\onlinecite{MelePRB2010}
do not coincide with those obtained by tight-binding
calculations~\cite{Pankratov4}. The value of the single-electron gap was evaluated for several commensurate
twist angles, see
Refs.~\onlinecite{Pankratov2,Pankratov4,MelePRB2010};
nonetheless, the generic dependence of the gap on $\theta$ was not discussed.

Here we report the results of tight-binding calculations of the band
structure of tBLG in a wide range of twist angles. To tackle the issue of
the large supercell size we use the Lanczos algorithm, which allows us to
calculate the low-energy single-electron spectrum of tBLG. We find that the
tBLG single-electron properties are qualitatively different for $\theta$
larger and smaller than the critical angle
$\theta_{\rm c}\approx 1.89^{\circ}$.
When
$\theta>\theta_{\rm c}$,
the low-energy spectrum  can be considered roughly as consisting of two
doubly degenerate Dirac cones located near two Dirac points in the
Brillouin zone of the superlattice. The Fermi velocity of Dirac electrons
is continuous function of the twist angle $\theta$, and it decreases
when $\theta$ decreases. This result is in agreement both with the
low-energy theories~\cite{dSPRL,dSPRB,PNAS} and the tight-binding
calculations, reported
elsewhere~\cite{Pankratov2,NanoLett,Morell}.

Under more scrutiny the spectrum reveals its fine structure: the
double-degeneracy of the single-electron bands is weakly lifted, and for
the momenta close to the Dirac points the dispersion deviates from massless
Dirac spectrum. The spectrum of the tBLG can be gapped or gapless depending
on the type of superstructure, but for any superstructure the band
splitting is non-zero. The maximum value of the band gap is estimated to be
$80$\,meV.
It corresponds to the twist angle
$\theta\cong 21.79^{\circ}$.

However, in contrast to the Fermi velocity, the band splitting is a
discontinuous function of the twist angle. It can change exponentially, even
for small variations of $\theta$. Such a feature makes it difficult to
predict the gap value for real samples, whose twist angles are always
known with some non-zero error. It is demonstrated that this sensitivity of
the gap to small variations of the angle is absent for a sample of
finite size. The relevance of this `smoothing' for experiment is discussed.

At the critical angle
$\theta_{\rm c}\cong1.89^{\circ}$,
the Fermi velocity vanishes, and for
$\theta<\theta_{\rm c}$
the cone-like structure of the low-energy bands becomes irrelevant.
Instead, the system has a finite density of states and a Fermi surface.
The Fermi surface changes smoothly as a function of $\theta$.

The presentation below is organized as follows. In
Sec.~\ref{Geometry}
we briefly discuss the geometry of the tBLG lattice. In
Sec.~\ref{TBM}
the tight-binding Hamiltonian is introduced. In
Sec.~\ref{sect::large_angle}
the case of large twist angles is discussed. Small $\theta$ are discussed
in
Sec.~\ref{sect::small_angle}.
Conclusions are presented in
Sec.~\ref{sect::conclusions}.

\section{Geometrical considerations}\label{Geometry}

Each graphene layer in the tBLG consists of two sublattices ($A1$, $B1$ in
the layer $1$, and $A2$, $B2$ in the layer $2$). The positions of the
carbon atoms in each sublattice in the
bottom layer $1$ are
\begin{equation}
\!\!\mathbf{r}_{\mathbf{n}}^{1A}\equiv\mathbf{r}_{\mathbf{n}}=n\mathbf{a}_1+m\mathbf{a}_2,\;\;\;
\mathbf{r}_{\mathbf{n}}^{1B}=\mathbf{r}_{\mathbf{n}}+\bm{\delta}_1,
\end{equation}
where $\mathbf{n}=\{n,m\}$ ($n$, $m$ are integers), $\bm{\delta}_1=(\mathbf{a}_1+\mathbf{a}_2)/3=a\{1/\sqrt{3},\,0\}$, and
$\mathbf{a}_{1,2}$ are basis vectors of the graphene elementary unit cell,
\begin{equation}
\!\!\mathbf{a}_1=a\{\sqrt{3},\,-1\}/2,\;\;\;\mathbf{a}_2=a\{\sqrt{3},\,1\}/2,
\end{equation}
with the lattice parameter $a=2.46$\,{\AA}. The distance between graphene layers
is $d=3.35$\,{\AA}. When the layers are not rotated
($\theta = 0$), the system is a perfect AB bilayer.

\begin{figure}[t]
\begin{center}
\includegraphics[width=0.75\columnwidth]{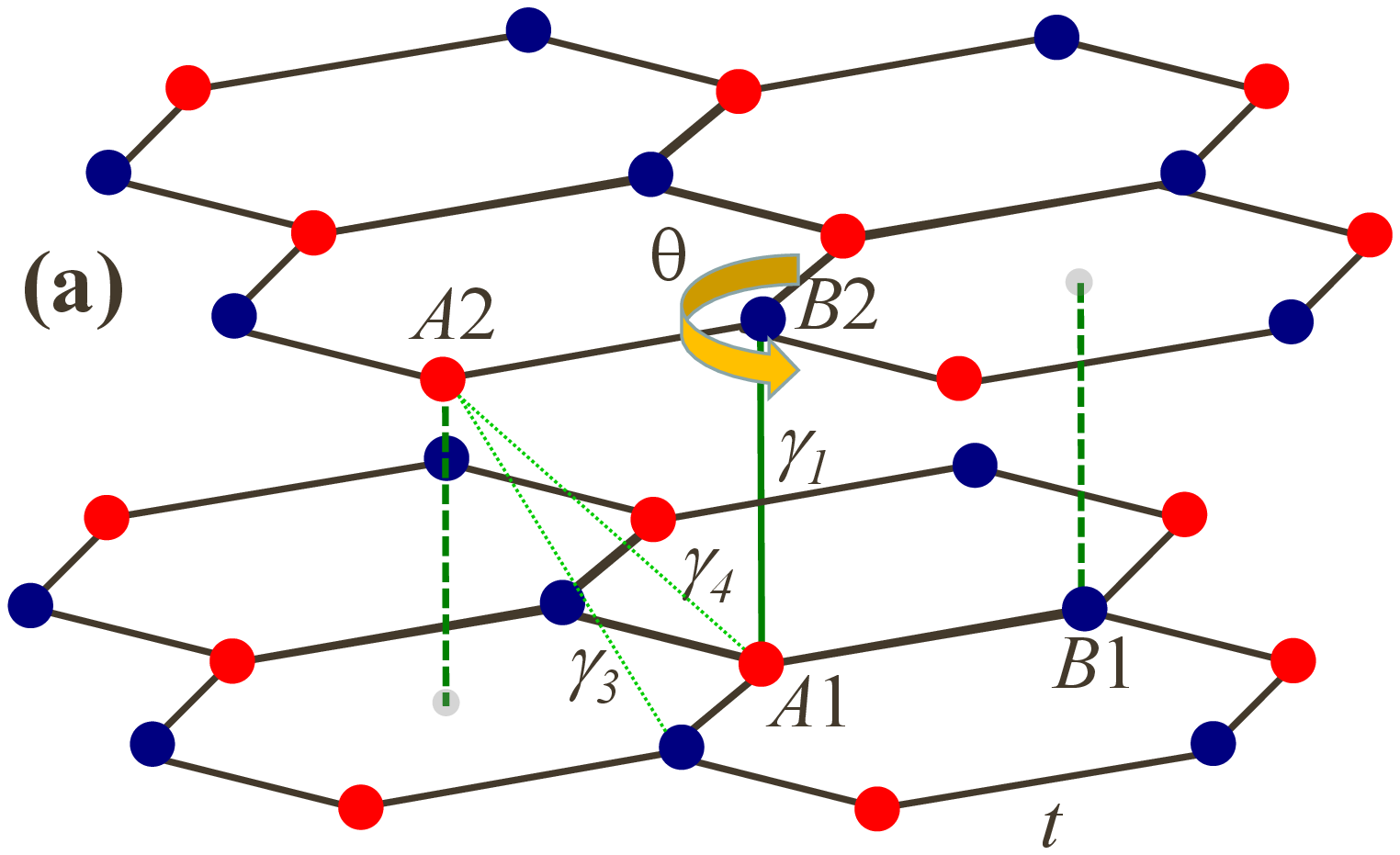}\\
\includegraphics[width=0.75\columnwidth]{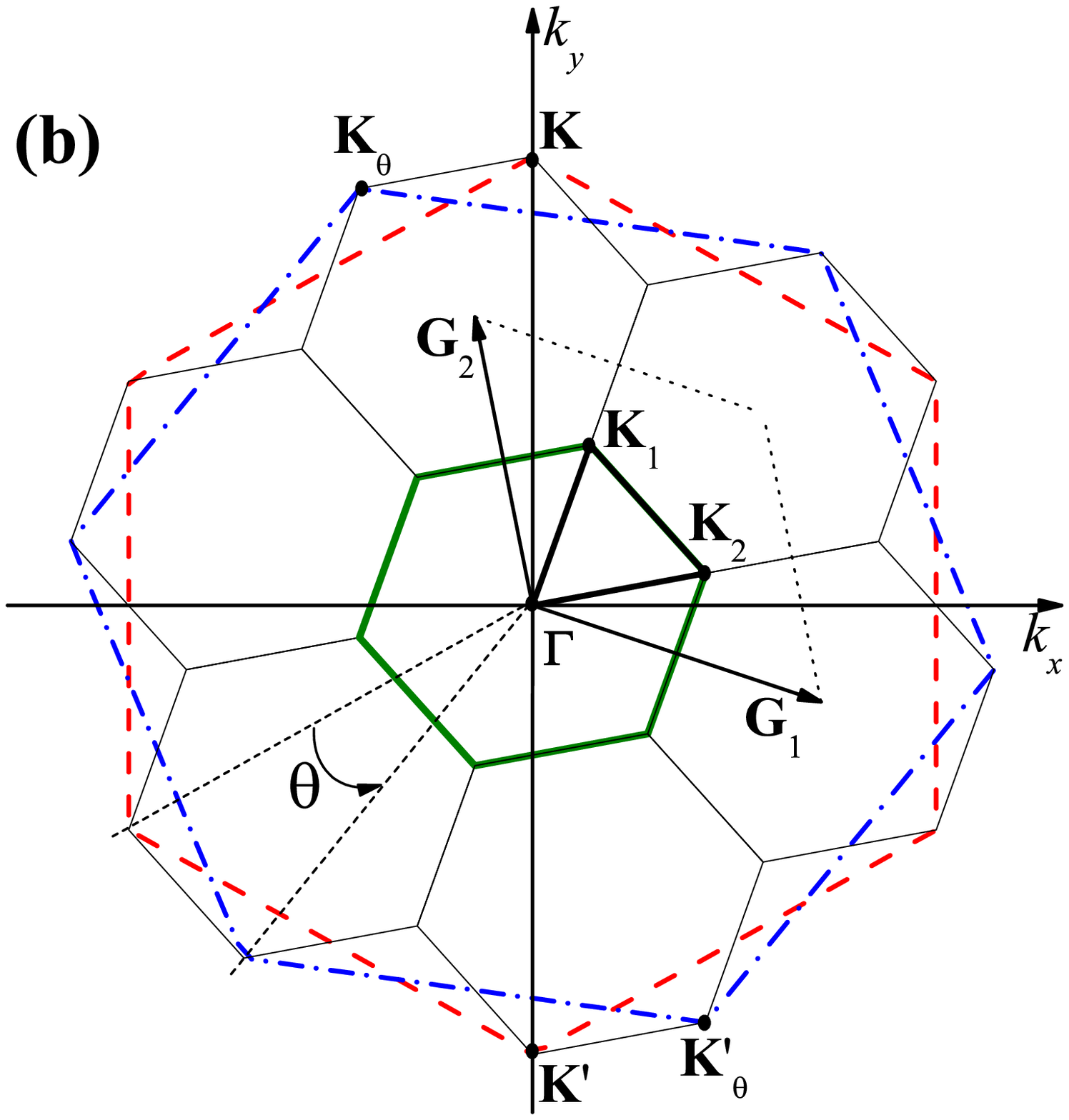}
\end{center}
\caption{(Color online)
(a) Structure of the AB-stacked graphene bilayer. A twisted graphene
bilayer is obtained by rotating the top layer by an angle $\theta$ around the
axis connecting sites $A1$ and $B2$; quantity $t$ is the in-plane
nearest-neighbor hopping, and
$\gamma_1$,
$\gamma_3$,
and
$\gamma_4$
are out-of-plane hopping amplitudes of the AB-stacked bilayer. These
$\gamma$s are used to fix the fitting parameters of the function
$t_{\bot}(\mathbf{r};\mathbf{r}')$
(see the text). In this paper we use
$\gamma_1 = 0.4$\;eV,
$\gamma_3 = 0.254$\;eV,
and
$\gamma_4 = 0.051$\;eV.
(b) The large hexagons show the Brillouin zones of individual layers: the
red dashed hexagon corresponds to the bottom layer, the blue dot-dashed
hexagon corresponds to the top layer for the twist angle
$\theta=21.787^{\circ}$
($m_0=1$,
$r=1$).
The first Brillouin zone of the bilayer is shown by the central (green) thick solid
hexagon. The next several Brillouin zones of the tBLG are depicted by the six surrounding
(black) thin solid hexagons.
The electronic spectra presented in Fig.~\ref{FigSpec}
are calculated along the path specified by the black triangle
${\bm\Gamma}\mathbf{K}_1\mathbf{K}_{2}$.
For the twisted bilayer, the Dirac point
$\mathbf{K}'$ ($\mathbf{K}'_{\theta}$)
is equivalent to the point
$\mathbf{K}_{\theta}$
($\mathbf{K}$)
if $r\neq3n$. When $r=3n$, $\mathbf{K}_{\theta}\sim\mathbf{K}$ and $\mathbf{K}'_{\theta}\sim\mathbf{K}'$ (see the text).
The tBLG Dirac points $\mathbf{K}_{1,2}$ are doubly degenerate: each of them is equivalent to one of two Dirac points of
each graphene layer. For the particular case of the $(1,1)$ superstructure, $\mathbf{K}_1\sim\mathbf{K}\sim\mathbf{K}'_{\theta}$
and $\mathbf{K}_2\sim\mathbf{K}'\sim\mathbf{K}_{\theta}$.}
\label{FigTBLG}
\end{figure}

Layer $2$ is rotated with respect to layer $1$ by the angle $\theta$ around the axis connecting the atoms $A1$ and $B2$ with $\mathbf{n}=0$ (see Fig.~\ref{FigTBLG}). In this case the atoms of the top layer have the positions
\begin{equation}
\!\!\mathbf{r}_{\mathbf{n}}^{2B}\equiv\mathbf{r}'_{\mathbf{n}}=n\mathbf{a}'_1+m\mathbf{a}'_2,\;\;\;
\mathbf{r}_{\mathbf{n}}^{2A}=\mathbf{r}'_{\mathbf{n}}-\bm{\delta}_2,
\end{equation}
where
\begin{equation}
\mathbf{a}'_{1,2}=\mathbf{a}_{1,2}\left(\cos\theta\mp\frac{\sin\theta}{\sqrt{3}}\,\right)\pm\mathbf{a}_{2,1}\frac{2\sin\theta}{\sqrt{3}}\,,
\end{equation}
and $\bm{\delta}_2=a\{\cos\theta,\sin\theta\}/\sqrt{3}$.

The structure of the tBLG is commensurate
if~\cite{dSPRL,dSPRB,Pankratov1,MeleReview}
\begin{equation}\label{theta}
\cos\theta=\frac{3m_0^2+3m_0r+r^2/2}{3m_0^2+3m_0r+r^2}\,,
\end{equation}
where
$m_0$
and $r$ are coprime positive integers. The superlattice vectors,
$\mathbf{R}_{1,2}$,
can be expressed via
$m_0$,
$r$, and the single-layer graphene lattice vectors,
$\mathbf{a}_{1,2}$.
These expressions are different when $r$ is either non-divisible or divisible
by three. In the former case, we have:
\begin{equation}
\begin{array}{rcl}
\mathbf{R}_1&=&m_0\mathbf{a}_1+(m_0+r)\mathbf{a}_2\\
\mathbf{R}_2&=&-(m_0+r)\mathbf{a}_1+(2m_0+r)\mathbf{a}_2
\end{array}\;(r\neq3n,\;n\in{\cal N})\,.
\end{equation}
For $r=3n$, the superlattice vectors become:
\begin{equation}
\begin{array}{rcl}
\mathbf{R}_1&=&(m_0+n)\mathbf{a}_1+n\mathbf{a}_2\\
\mathbf{R}_2&=&-n\mathbf{a}_1+(m_0+2n)\mathbf{a}_2
\end{array}\;(r=3n,\;n\in{\cal N})\,.
\end{equation}
The number of sites in each supercell is:
\begin{equation}
\label{Nsc}
N(m_0,r)=\left\{\begin{array}{l}
4(3m_0^2+3m_0r+r^2),\;\text{if}\;r\neq3n,\\
4(m_0^2+m_0r+r^2/3),\;\text{if}\;r=3n\,.
\end{array}\right.
\end{equation}
The linear size of the superlattice cell (SC) is
\begin{equation}
L_{\text{sc}}\equiv|\mathbf{R}_{1,2}|=a\sqrt{N}/2\,.
\end{equation}

Besides
$L_{\text{sc}}$,
the tBLG has another characteristic length scale. The rotation of one
graphene layer with respect to another one leads to the appearance of
Moir\'{e} patterns, which manifest themselves as alternating bright and dark
regions in STM
images~\cite{STM_LL,STMNanoLett,STM_VHS,STM_VHS2}.
The Moir\'{e} period $L$, is defined as the distance between centers of two
neighboring bright (or dark) regions. It is related to the twist angle
according to the following formula
\begin{equation}
\label{MoireP}
L=\frac{a}{2\sin(\theta/2)}\,.
\end{equation}
The Moir\'{e} pattern and the superstructure are two complementary
concepts used to describe tBLG. The Moir\'{e} pattern depends smoothly on the
twist angle, see, e.g.,
Eq.~(\ref{MoireP}).
The pattern can be easily detected experimentally. However, working with
the Moir\'{e} theoretically may be challenging due to the fact that the
pattern is strictly periodic only for a very limited discrete set of angles. For a generic value of $\theta$, different Moir\'{e} cells in a
pattern may look alike, but they are not exactly identical.

The superstructure, which is a periodic lattice of supercells, does not
suffer from this shortcoming. Unfortunately, it has its own deficiencies.
Namely, the superstructure is only defined for commensurate angles $\theta$. The period
$L_{\text{sc}}$
is not a smooth function of $\theta$: two commensurate $\theta$ and
$\theta'$,
$\theta \approx \theta'$,
may correspond to two very dissimilar
$L_{\text{sc}}$.
As we will see below, such sensitivity to the twist angle may, in some
situations, require additional efforts in interpreting theoretical results.

One can easily demonstrate that the superstructure coincides with the
Moir\'{e} pattern only when
$r=1$.
For other superstructures,
$L_{\text{sc}}$
is greater than $L$. The supercells of these structures contain
$r^2$
(if
$r\neq3n$)
or
$r^2/3$
(if
$r=3n$)
Moir\'{e} cells, and the arrangements of atoms inside these Moir\'{e} cells
are slightly different from each other. This means, in
particular, that the structures with $r>1$ can be considered as almost
periodic repetitions of structures with $r=1$, as it was shown in
Ref.~\onlinecite{dSPRB}.

The basis vectors of the reciprocal superlattice can be written as
\begin{eqnarray}
\mathbf{G}_1&=&\frac{(2m_0+r)\mathbf{b}_1+(m_0+r)\mathbf{b}_2}{3m_0^2+3m_0r+r^2}\,,\nonumber\\
\mathbf{G}_2&=&\frac{-(m_0+r)\mathbf{b}_1+m_0\mathbf{b}_2}{3m_0^2+3m_0r+r^2}\,,
\end{eqnarray}
if $r\neq3n$, or
\begin{eqnarray}
\mathbf{G}_1&=&\frac{(m_0+2n)\mathbf{b}_1+n\mathbf{b}_2}{m_0^2+m_0r+r^2/3}\,,\nonumber\\
\mathbf{G}_2&=&\frac{-n\mathbf{b}_1+(m_0+n)\mathbf{b}_2}{m_0^2+m_0r+r^2/3}\,,
\end{eqnarray}
if $r=3n$, where
\begin{equation}
\mathbf{b}_1=2\pi\{1/\sqrt{3},\,-1\}/a,\;\;\;\mathbf{b}_2=2\pi\{1/\sqrt{3},\,1\}/a\,,
\end{equation}
are the reciprocal lattice vectors of the single-layer graphene. The first
Brillouin zone of the superlattice has the shape of a hexagon with sides
$|\mathbf{G}_{2}-\mathbf{G}_{1}|/3$.
In the particular case of $r=1$, this side is equal to $\Delta
K=|\mathbf{K}_{\theta}-\mathbf{K}|$,
where
\begin{eqnarray}
\mathbf{K}=\frac{4\pi}{3}\{0,\,1\}\;\;\;\text{and}\;\;\;
\mathbf{K}_{\theta}=\frac{4\pi}{3}\{-\sin\theta,\,\cos\theta\}
\end{eqnarray}
are the Dirac
points of the bottom and top layers, respectively.

As known from basic graphene tight-binding physics, in addition to the
Dirac cone at the $\mathbf{K}$ point, the bottom layer of the tBLG has another cone of opposite chirality at
$\mathbf{K}'=-\mathbf{K}$.
Likewise, the top layer has its second cone at
$\mathbf{K}'_{\theta}=-\mathbf{K}_{\theta}$.
It is important to determine where these two cones are located in the
Brillouin zone of the superstructure. To find this out we express their
co-ordinates in terms of reciprocal superlattice vectors. For
$r\neq3n$
we have
\begin{eqnarray}
\mathbf{K}&=&
-\mathbf{K}'=
m_0\mathbf{G}_2+\frac{r}{3}\left(\mathbf{G}_1+2\mathbf{G}_2\right),
\nonumber\\
\mathbf{K}_{\theta}&=&
-\mathbf{K}'_{\theta}=
m_0\mathbf{G}_2+
\frac{r}{3}\left(\mathbf{G}_2-\mathbf{G}_1\right)\,,\label{Kn3n}
\end{eqnarray}
while for $r=3n$ we obtain
\begin{eqnarray}
\mathbf{K}&=&-\mathbf{K}'=\frac{r}{3}\mathbf{G}_2+\frac{m_0}{3}\left(\mathbf{G}_2-\mathbf{G}_1\right),\nonumber\\
\mathbf{K}_{\theta}&=&-\mathbf{K}'_{\theta}=-\frac{r}{3}\mathbf{G}_1+\frac{m_0}{3}\left(\mathbf{G}_2-\mathbf{G}_1\right)\,.\label{K3n}
\end{eqnarray}
It follows from these formulas that, if
$r\neq3n$,
point
$\mathbf{K}'$
is equivalent to
$\mathbf{K}_{\theta}$,
and
$\mathbf{K}$
is equivalent to
$\mathbf{K}'_{\theta}$:
for such a value of $r$ the difference
$\mathbf{K}'-\mathbf{K}_{\theta}$
is a reciprocal vector of the superlattice. When
$r=3n$, the equivalent Dirac points are:
$\mathbf{K}\sim\mathbf{K}_{\theta}$ and
$\mathbf{K}'\sim\mathbf{K}'_{\theta}$. Thus, for any commensurate angle we
have two doubly-degenerate non-equivalent Dirac points of the tBLG. It follows from Eqs.~\eqref{Kn3n} and~\eqref{K3n} that inside
the reciprocal cell of the superlattice, the two non-equivalent tBLG Dirac points are located at:
\begin{equation}\label{K12}
\mathbf{K}_1=\frac{\mathbf{G}_1+2\mathbf{G}_2}{3}\,,\;\;\mathbf{K}_2=\frac{2\mathbf{G}_1+\mathbf{G}_2}{3}\,,
\end{equation}
for any superstructure. As we will show below, this double degeneracy affects the
electronic structure of the tBLG, leading to band splitting and band gap
formation.

\section{Tight-binding Hamiltonian}\label{TBM}

It is convenient to enumerate the sites in the sublattice in each layer using two
integer-valued vectors
$\mathbf{j}=\{i,j\}$
and
$\mathbf{n}=\{n,m\}$,
where
$\mathbf{j}$
labels the position of the supercell in the lattice, while
$\mathbf{n}$
enumerates the sites inside the supercell. Then, we can write down the
tight-binding Hamiltonian of the tBLG in the form
\begin{eqnarray}
H\!\!&=&\!\!-t\!\!\!\sum_{\langle\mathbf{in},\mathbf{jm}\rangle\atop s\sigma}
\!\!\!\!\left(d^{\dag}_{s\mathbf{in}A\sigma}d^{\phantom{\dag}}_{s\mathbf{jm}B\sigma}+H.c.\right)+\label{H}\\
\!\!&&\!\!\!\!\sum_{{\mathbf{in},\mathbf{jm}\atop\alpha\beta\sigma}}\left[
t_{\bot}(\mathbf{R}_{\mathbf{i}}+\mathbf{r}_{\mathbf{n}}^{1\alpha};\mathbf{R}_{\mathbf{j}}+\mathbf{r}_{\mathbf{m}}^{2\beta})
d^{\dag}_{1\mathbf{in}\alpha\sigma}d^{\phantom{\dag}}_{2\mathbf{jm}\beta\sigma}+H.c.\right]\!,\nonumber
\end{eqnarray}
where
$\mathbf{R}_{\mathbf{j}}=i\mathbf{R}_1+j\mathbf{R}_2$,
the symbol
$\langle\dots\rangle$
stands for summation over the nearest neighbors within the same layer,
$d^{\dag}_{s\mathbf{jn}\alpha\sigma}$
and
$d^{\phantom{\dag}}_{s\mathbf{jn}\alpha\sigma}$
are the creation and annihilation operators of an electron with the spin
projection $\sigma$ in the layer
$s$\,($=1,2$)
on the sublattice
$\alpha$\,($=A,B$)
in the supercell
$\mathbf{j}$
in the position
$\mathbf{n}$
(the position of this site is
$\mathbf{R}_{\mathbf{j}}+\mathbf{r}_{\mathbf{n}}^{s\alpha}$).
The first term describes the in-plane nearest-neighbor hopping with amplitude
$t=2.57$\,eV.
The second term describes the interlayer hopping, with
$t_{\bot}(\mathbf{r};\mathbf{r}')$
being the hopping amplitude between sites in the positions
$\mathbf{r}$
and
$\mathbf{r}'$.

This Hamiltonian~\eqref{H} is invariant with respect to translations by the
superstructure vectors
$\mathbf{R}_{1,2}$.
Performing the Fourier transform
$d_{s\mathbf{kn}\alpha\sigma}=
{\cal N}_{sc}^{-1/2}
\sum_{\mathbf{j}}e^{-i\mathbf{k}\mathbf{R}_{\mathbf{j}}}
d_{s\mathbf{j}\mathbf{n}\alpha\sigma}$,
where
${\cal N}_{sc}$
is the number of supercells in the bilayer, and using the relation
$t_{\bot}(\mathbf{R}_{\mathbf{j}}+\mathbf{r};
\mathbf{R}_{\mathbf{j}}+\mathbf{r}')=t_{\bot}(\mathbf{r};\mathbf{r}')$,
we obtain
\begin{eqnarray}
\label{Hinit}
H&=&\sum_{\mathbf{k}\mathbf{n}\,\mathbf{m}\atop s\sigma}\!
\left[t^{s}_{\mathbf{nm}}(\mathbf{k})
d^{\dag}_{s\mathbf{k}\mathbf{n}A\sigma}
d^{\phantom{\dag}}_{s\mathbf{k}
\mathbf{m}B\sigma}+H.c.\right]+
\nonumber\\
&&\sum_{{\mathbf{k}\mathbf{n}\,\mathbf{m}\atop\alpha\beta\sigma}}
\!\left[t_{\bot\mathbf{nm}}^{\alpha\beta}(\mathbf{k})
d^{\dag}_{1\mathbf{k}\mathbf{n}\alpha\sigma}
d^{\phantom{\dag}}_{2\mathbf{k}\mathbf{m}\beta\sigma}\!+\!H.c.\right]\!,
\label{Hk}
\end{eqnarray}
where
$\mathbf{k}$
runs over the first Brillouin zone of the superlattice. In
Eq.~\eqref{Hinit}
\begin{eqnarray}
t^{s}_{\mathbf{nm}}(\mathbf{k})&=&-t\sum_{\mathbf{j}\bm{\delta}}
e^{-i\mathbf{k}\mathbf{R_{j}}}
\delta_{\mathbf{N}_{\mathbf{j}}+\mathbf{n},\,\mathbf{m}-\bm{\delta}}\,,\label{tin}\\
t_{\bot\mathbf{nm}}^{\alpha\beta}(\mathbf{k})&=&
\sum_{\mathbf{j}}e^{-i\mathbf{kR_{j}}}t_{\bot}(\mathbf{R_j}+\mathbf{r}_{\mathbf{n}}^{1\alpha};\mathbf{r}_{\mathbf{m}}^{2\beta})\,,\label{tout}
\end{eqnarray}
the vector
$\bm{\delta}$
takes the values $\{0,0\}$, $\{1,0\}$, $\{0,1\}$, and
$$
\mathbf{N}_{\mathbf{j}}=\{m_0i-(m_0+r)j,(m_0+r)i+(2m_0+r)j\}\,.
$$

We use the approach proposed in Ref.~\onlinecite{Tang} to calculate the interlayer hopping amplitudes.
The main premise of this approach is that $t_{\bot}(\mathbf{r};\mathbf{r}')$
depends not only on the relative positions of the initial and final carbon
atoms, but also on the positions of other atoms in the bilayer via the
screening function $S(\mathbf{r};\mathbf{r}')$
[see Eq. (2) in Ref.~\onlinecite{Tang}];
the closer some of the neighboring atoms are to the line connecting the sites
$\mathbf{r}$ and $\mathbf{r}'$,
the stronger is the screening. The inclusion of the screening is very
important. Otherwise, the longer-range hopping amplitudes in the usual
Slonczewski-Weiss-McClure (SWM)
scheme~\cite{McClSlW,Mendez,Mucha,Dress}
cannot be correctly reproduced. Without screening, the next-nearest-neighbor interlayer hopping amplitudes of the AB bilayer,
$\gamma_3$
and
$\gamma_4$
[see
Fig~\ref{FigTBLG}(a)],
become identical. This conclusion is at odds with the SWM scheme, where
these amplitudes differ by about an order of magnitude.

Following Ref.~\onlinecite{Tang} we write the hopping amplitude in the form
\begin{eqnarray}
&&t_{\bot}(\mathbf{r};\mathbf{r}')=
\cos^2\!\alpha\,V_{\sigma}(\mathbf{r};\mathbf{r}')
+\sin^2\!\alpha\,V_{\pi}(\mathbf{r};\mathbf{r}')\,,\nonumber\\
&&\cos\alpha=\frac{d}{\sqrt{d^2+(\mathbf{r}-\mathbf{r}')^2}}\,,
\end{eqnarray}
where the `Slater-Koster' functions $V_{\sigma}$ and $V_{\pi}$ contain the factor $[1-S(\mathbf{r};\mathbf{r}')]$ (exact expressions for $V_{\sigma}$ and $V_{\pi}$ are given by Eq.~(1) in Ref.~\onlinecite{Tang}). We found that the contribution to $t_{\bot}$ from $V_{\pi}$ is negligible, in agreement with Refs.~\onlinecite{Pankratov3,dSPRL,dSPRB}. Due to screening, the function $t_{\bot}(\mathbf{r};\mathbf{r}')$ decays very quickly when $|\mathbf{r}-\mathbf{r}'|>a$. The functions $V_{\sigma}(\mathbf{r};\mathbf{r}')$ and $S(\mathbf{r};\mathbf{r}')$ in Ref.~\onlinecite{Tang} depend on seven fitting parameters ($\alpha_{1,2,3,4}$ and $\beta_{1,2,3}$).
However, the values found in Ref.~\onlinecite{Tang} cannot be directly applied to bilayer graphene~\cite{Pankratov3}. Instead, we use the following estimates for the fitting
constants~\cite{note1}: $\alpha_1=6.715$, $\alpha_2=0.762$, $\alpha_3=0.179$, $\alpha_4=1.411$, $\beta_1=6.811$, $\beta_2=0.01$, and $\beta_3=19.176$ (c.f., with the third line of Table~1 of Ref.~\onlinecite{Tang}). With these coefficients we reproduce the well-known SWM hopping amplitudes $\gamma_{1,3,4}$ in the AB bilayer (see Fig.~\ref{FigTBLG} for the definitions of $\gamma_{1,3,4}$). A similar approach was used in Ref.~\onlinecite{Pankratov3}, but the authors obtained different fitting parameters because they used another optimization procedure.

We now introduce the $N$-component operator
\begin{equation}
\Psi^{\dag}_{\mathbf{k}\sigma}=\{d^{\dag}_{1\mathbf{k}\mathbf{n}A\sigma},d^{\dag}_{1\mathbf{k}\mathbf{n}B\sigma},
d^{\dag}_{2\mathbf{k}\mathbf{n}A\sigma},d^{\dag}_{2\mathbf{k}\mathbf{n}B\sigma}\}
\end{equation}
and rewrite Eq.~\eqref{Hk}
in the form $H=\sum_{\mathbf{k}\sigma}\Psi^{\dag}_{\mathbf{k}\sigma}\hat{H}_{\mathbf{k}}\Psi^{\phantom{\dag}}_{\mathbf{k}\sigma}$, where $\hat{H}_{\mathbf{k}}$ is the $N\times N$ matrix
\begin{equation}
\hat{H}_{\mathbf{k}}=\left(\begin{array}{cccc}0&\hat{t}_{\mathbf{k}}^1&\hat{t}_{\bot\mathbf{k}}^{11}&\hat{t}_{\bot\mathbf{k}}^{12}\\
\hat{t}_{\mathbf{k}}^{1\dag}&0&\hat{t}_{\bot\mathbf{k}}^{21}&\hat{t}_{\bot\mathbf{k}}^{22}\\
\hat{t}_{\bot\mathbf{k}}^{11\dag}&\hat{t}_{\bot\mathbf{k}}^{21\dag}&0&\hat{t}_{\mathbf{k}}^2\\
\hat{t}_{\bot\mathbf{k}}^{12\dag}&\hat{t}_{\bot\mathbf{k}}^{22\dag}&\hat{t}_{\mathbf{k}}^{2\dag}&0
\end{array}\right).
\end{equation}
Here, the matrices $\hat{t}_{\mathbf{k}}^{s}$ and $\hat{t}_{\bot\mathbf{k}}^{\alpha\beta}$ are constructed from
$t^{s}_{\mathbf{nm}}(\mathbf{k})$ and $t_{\bot\mathbf{nm}}^{\alpha\beta}(\mathbf{k})$ according to Eqs.~\eqref{tin} and~\eqref{tout}.

The energy spectrum of the tBLG consists of $N$ bands
$E_{\mathbf{k}}^{(i)}$ ($i=1,\dots,N$).
We are interested here in the spectrum near the Fermi level
$\mu$ at half-filling. The chemical potential $\mu$ is non-zero
due to the violation of the particle-hole symmetry of our tight-binding
Hamiltonian, and it has to be found with the help of the charge-neutrality
requirement. For a rough estimate of $\mu$ we use the formula relating the
number of electrons per site $n_e$ and chemical potential $\mu$, which does
not require the full diagonalization of the matrix
$\hat{H}_{\mathbf{k}}$:
\begin{equation}\label{nvsmu}
n_e(\mu)=\frac{2T}{N}\!\!
\sum_{i\omega}\!\!\int\!\!\frac{d^2\mathbf{k}}{\upsilon_{\text{BZ}}}e^{i\omega0^{+}}\!\!\Tr\!\!\left[\frac{1}{i\omega+\mu-\hat{H}_{\mathbf{k}}}\right]=1\,,
\end{equation}
where $\upsilon_{\text{BZ}}$ is the area of the first Brillouin zone of the
superstructure, the summation is performed over the Matsubara frequencies,
and $T$ is the temperature, which we choose low enough ($T=0.1\gamma_1$).
Estimates of $\mu$ according to
Eq.~\eqref{nvsmu}
show that
$|\mu|$
is very small for any twist angle, and only $n_0$ bands with the smallest
absolute values,
$\bar{E}_{\mathbf{k}}^{(\nu)}=E_{\mathbf{k}}^{(i_0+\nu)}$
($i_0=(N-n_0)/2-2$, $\nu=1,\dots,n_0$)
can cross the Fermi level. Analysis shows that when $\theta>\theta_c$ the
number of low-energy bands is $n_0=4$. For $\theta<\theta_c$, we have
$n_0=4r^2$, if $r\neq3n$, or $n_0=4r^2/3$ otherwise. More precise value
of the chemical potential $\mu$ is found from the usual charge-neutrality
relation:
\begin{equation}
2\sum_{\nu}\int\!\!\frac{d^2\mathbf{k}}{\upsilon_{\text{BZ}}}\;
\Theta\!\left(\mu-\bar{E}_{\mathbf{k}}^{(\nu)}\right)=n_0\,.
\end{equation}

\begin{figure*}
\includegraphics[width=0.48\textwidth]{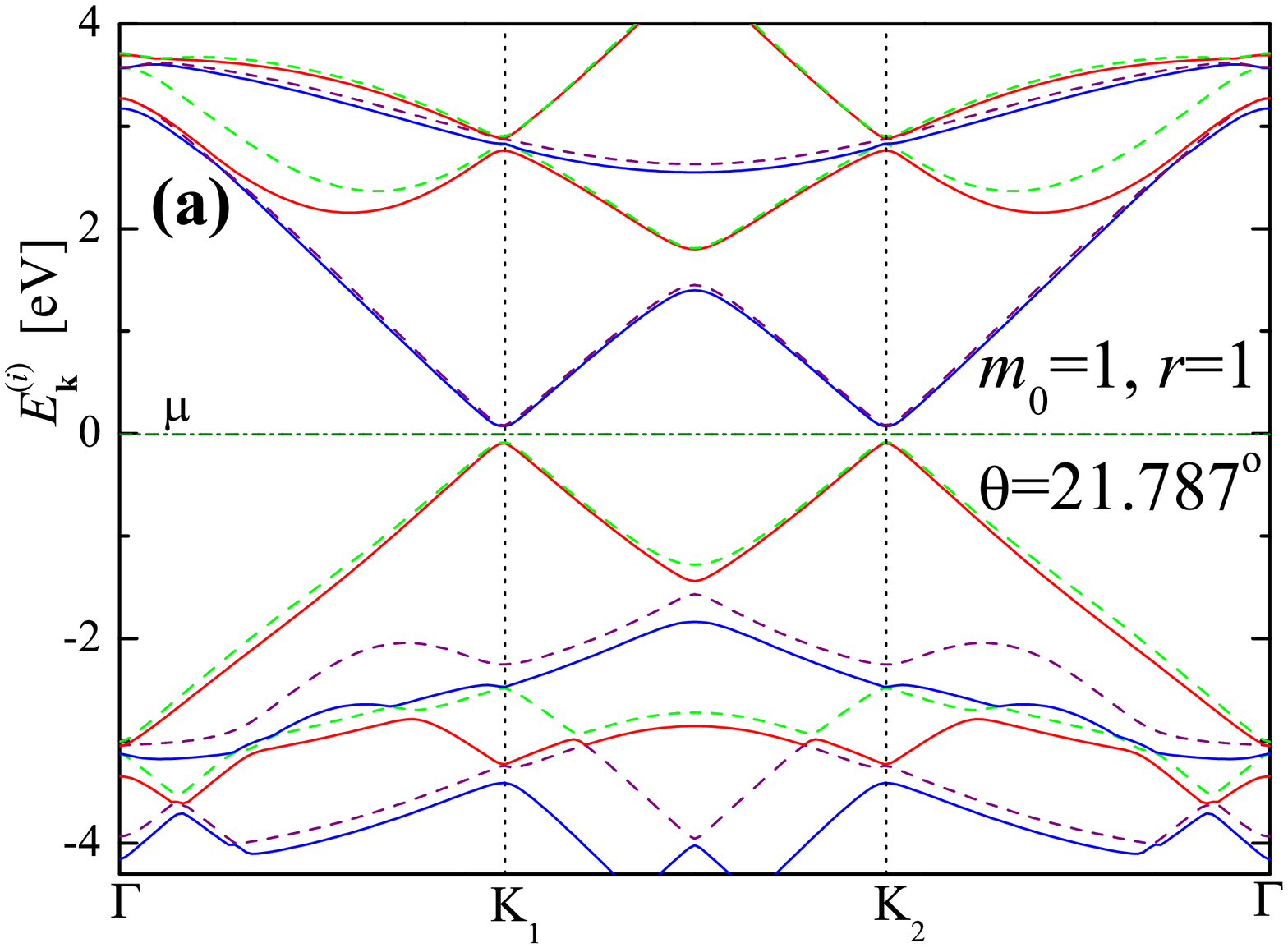}\hspace{5mm}
\includegraphics[width=0.48\textwidth]{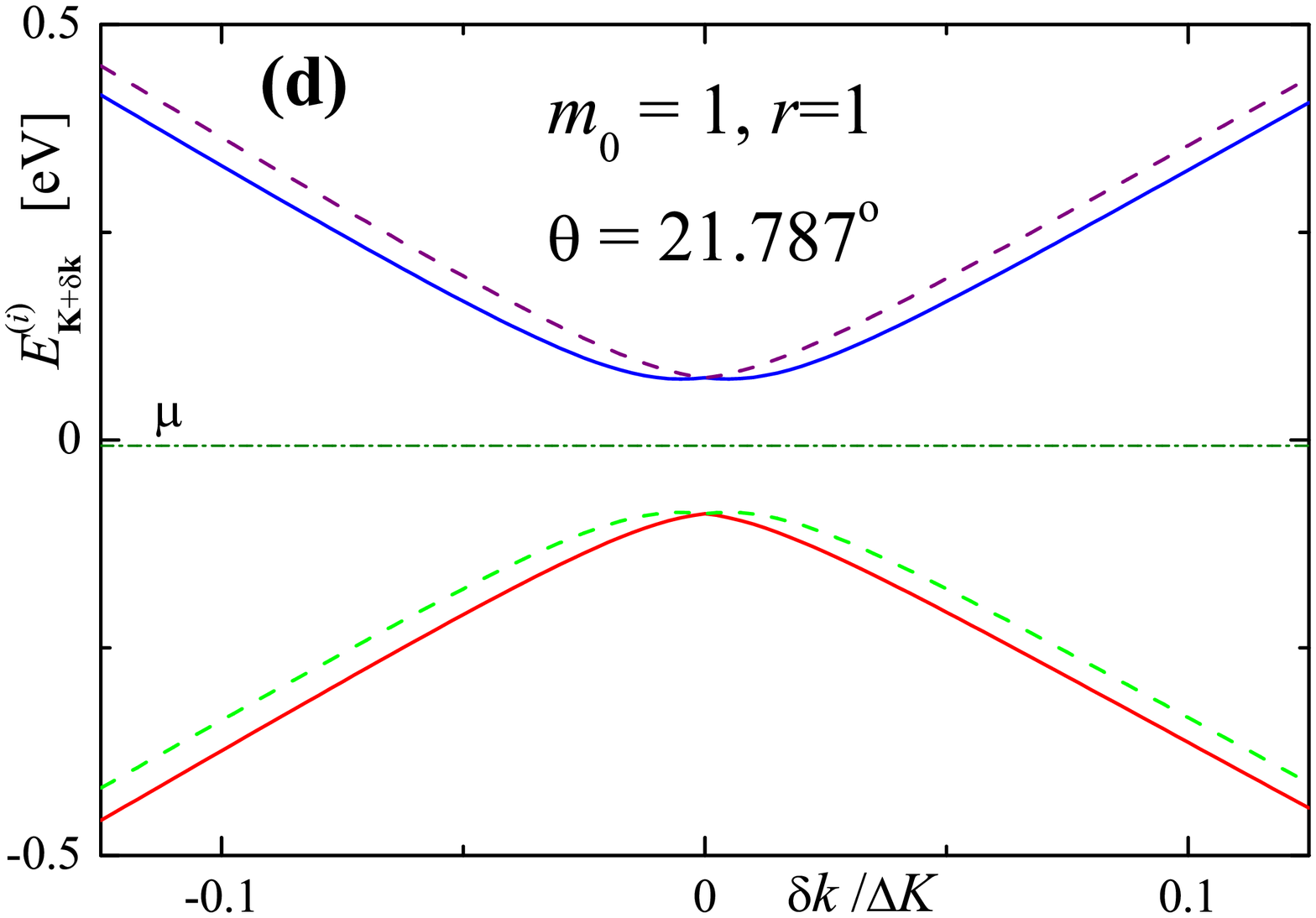}\\
\includegraphics[width=0.48\textwidth]{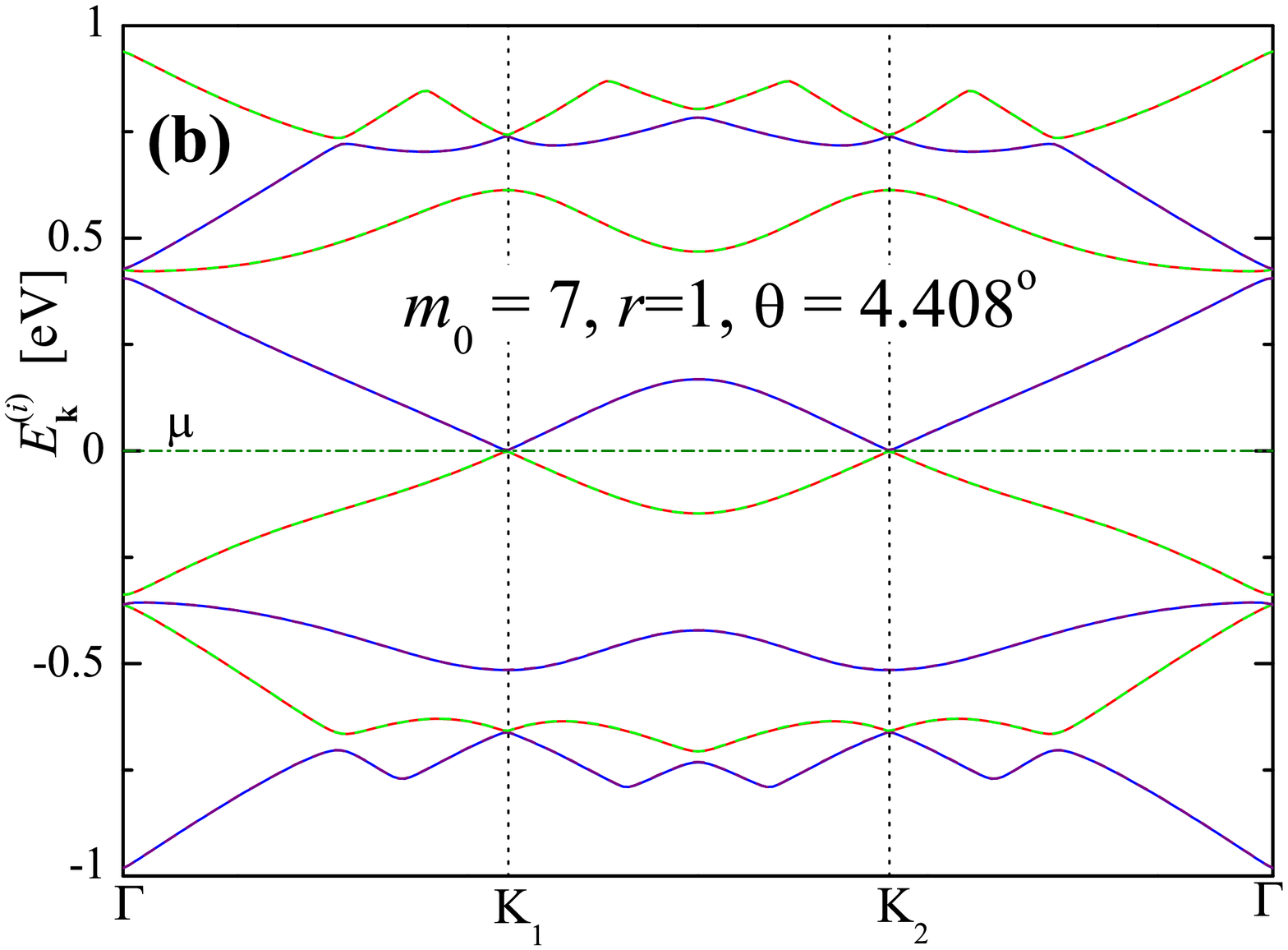}\hspace{5mm}
\includegraphics[width=0.48\textwidth]{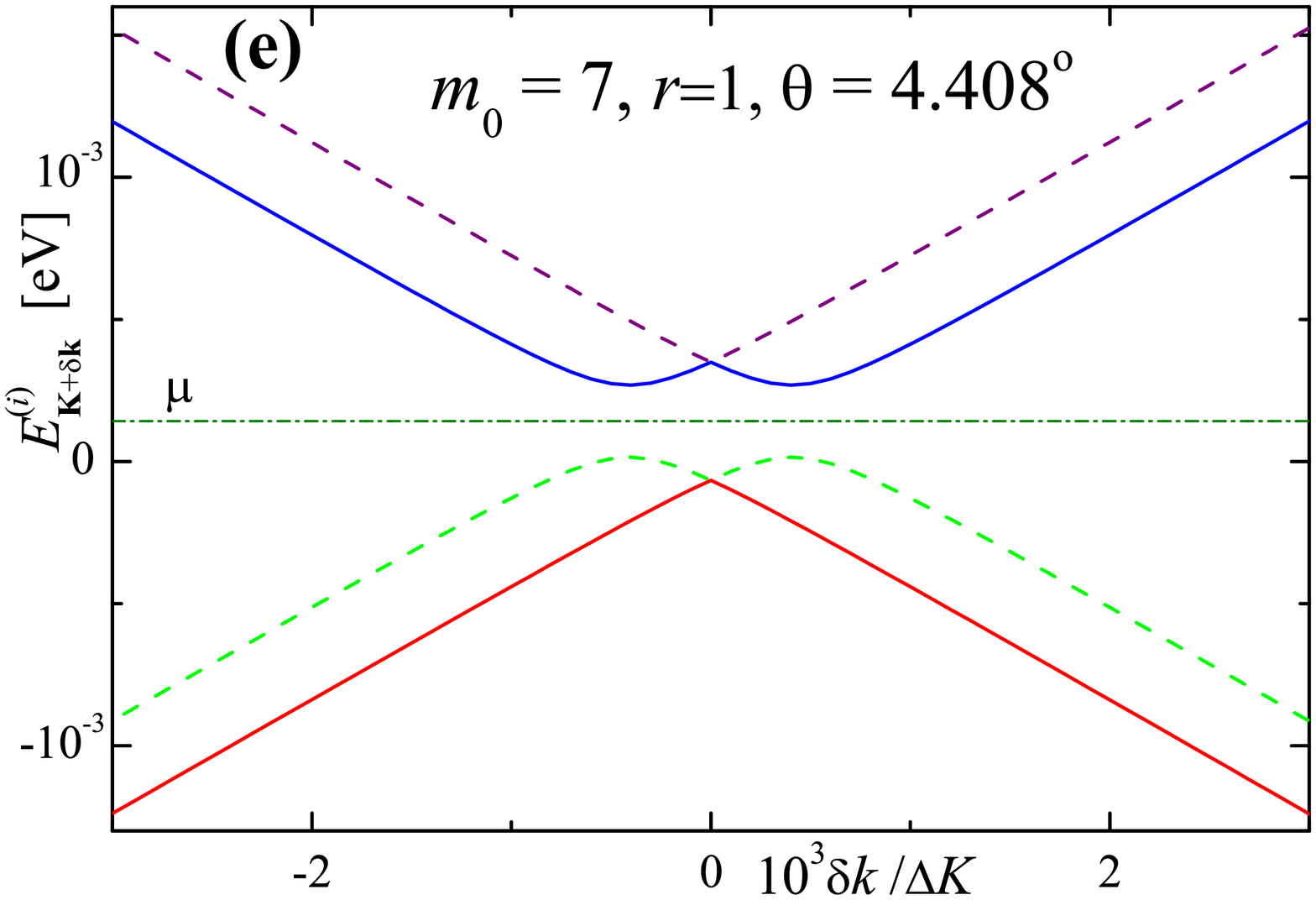}\\
\includegraphics[width=0.48\textwidth]{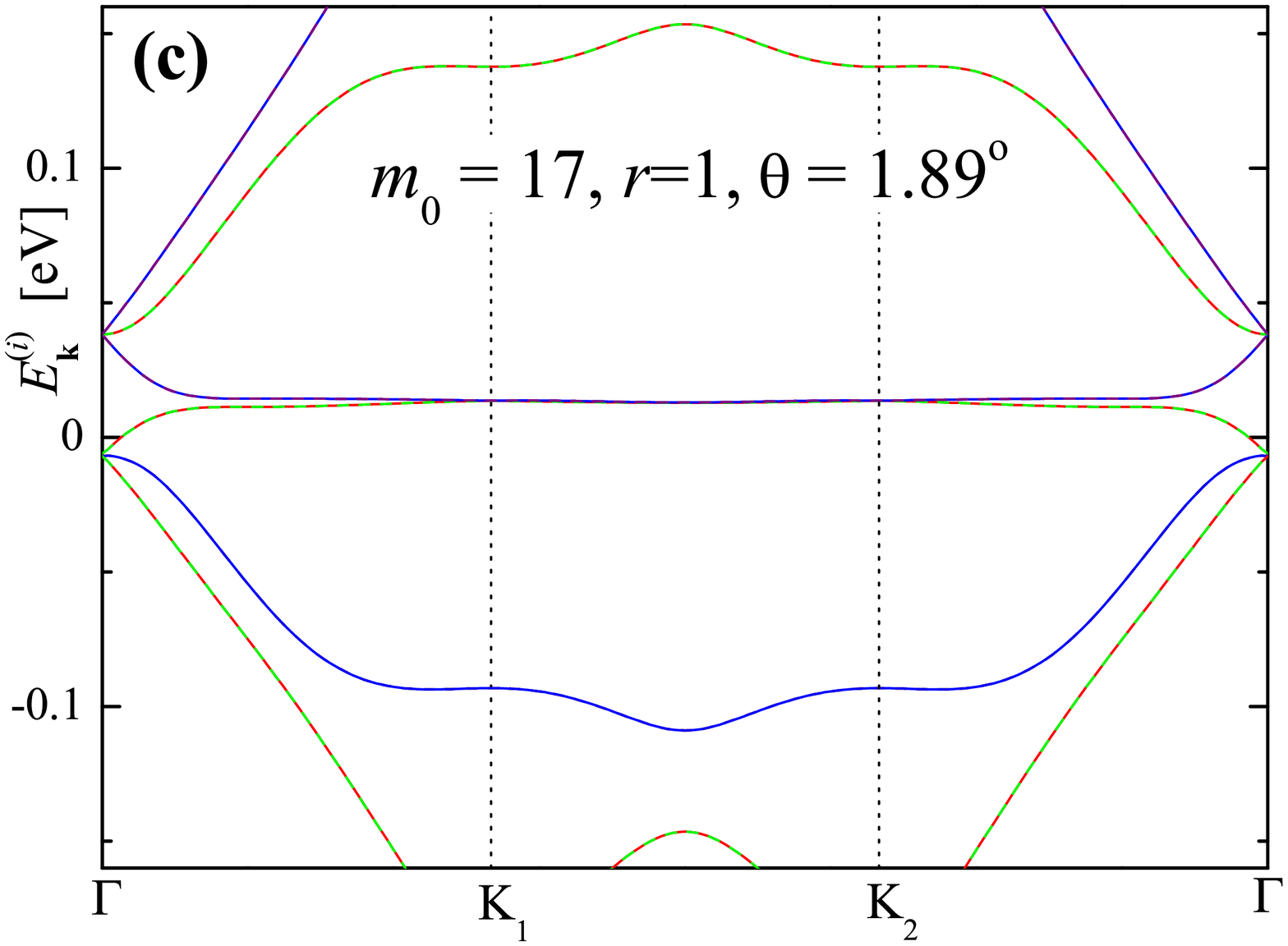}\hspace{6mm}
\includegraphics[width=0.47\textwidth]{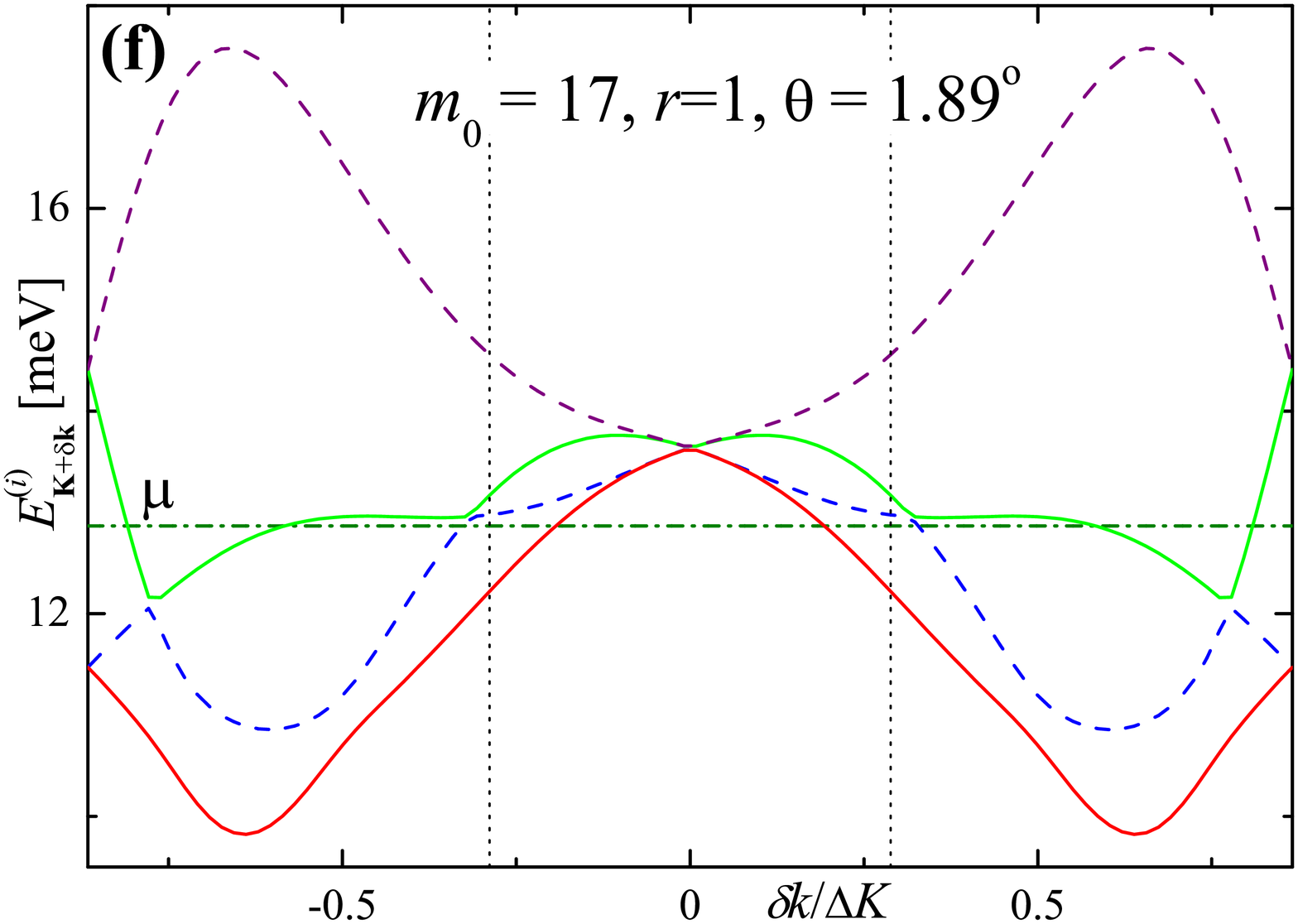}
\caption{(Color online) (a)--(c) The spectra of the twisted bilayer
graphene calculated for three different twist angles $\theta$ along the
path
${\bm\Gamma}\mathbf{K}_1\mathbf{K}_2$
shown in
Fig.~\ref{FigTBLG}(b).
The spectrum shown in panel~(a) corresponds to
$\theta\approx21.787^\circ$.
It demonstrates a significant gap. The detailed behavior of this spectrum
near the Dirac point is shown in panel~(d)
[the dispersion curves shown in panel~(d) and~(e) are calculated along the line
parallel to vector
$\Delta\mathbf{K}=\mathbf{K}_2-\mathbf{K}_1$].
Two almost-degenerate bands approaching the Fermi level $\mu$ from above
and two almost-degenerate bands approaching it from below are clearly seen.
For a much smaller angle $\theta\approx 4.408^\circ$, panel~(b), the gap is much smaller, but still present, see panel~(e).
Panels~(c) and~(f) correspond to
$\theta\approx1.89^\circ$.
The spectrum is gapless and three bands cross the Fermi energy
forming the Fermi surface.
The low-energy dispersion shown in panel~(f) is calculated along the line
passing through the Dirac point
$\mathbf{K}_2$
perpendicular to the vector
$\Delta\mathbf{K}$
[the dot-dashed line in Fig.~\ref{FigFSr1}(a)].
In panels (d)--(f)
the Dirac point corresponds to
${\bm\delta}\mathbf{k}=0$.
}\label{FigSpec}
\end{figure*}

The distant hopping amplitudes $t_{\bot}(\mathbf{r};\mathbf{r}')$ turn out to be negligible, and the matrix
$\hat{H}_{\mathbf{k}}$ is very sparse; that is, the number of non-zero
elements in $\hat{H}_{\mathbf{k}}$ is proportional to $N$. This allows us
to use the Lanczos algorithm to calculate the eigenvalues closest to zero
energy~\cite{arpackurls}. Then we calculate the spectra along the contour
in $\mathbf{k}$ space shown in Fig.~\ref{FigTBLG}(b) for a set of
superstructures with $(m_0,r)$ varying in a broad range.

\begin{figure}
\centering
\includegraphics[width=1\columnwidth]{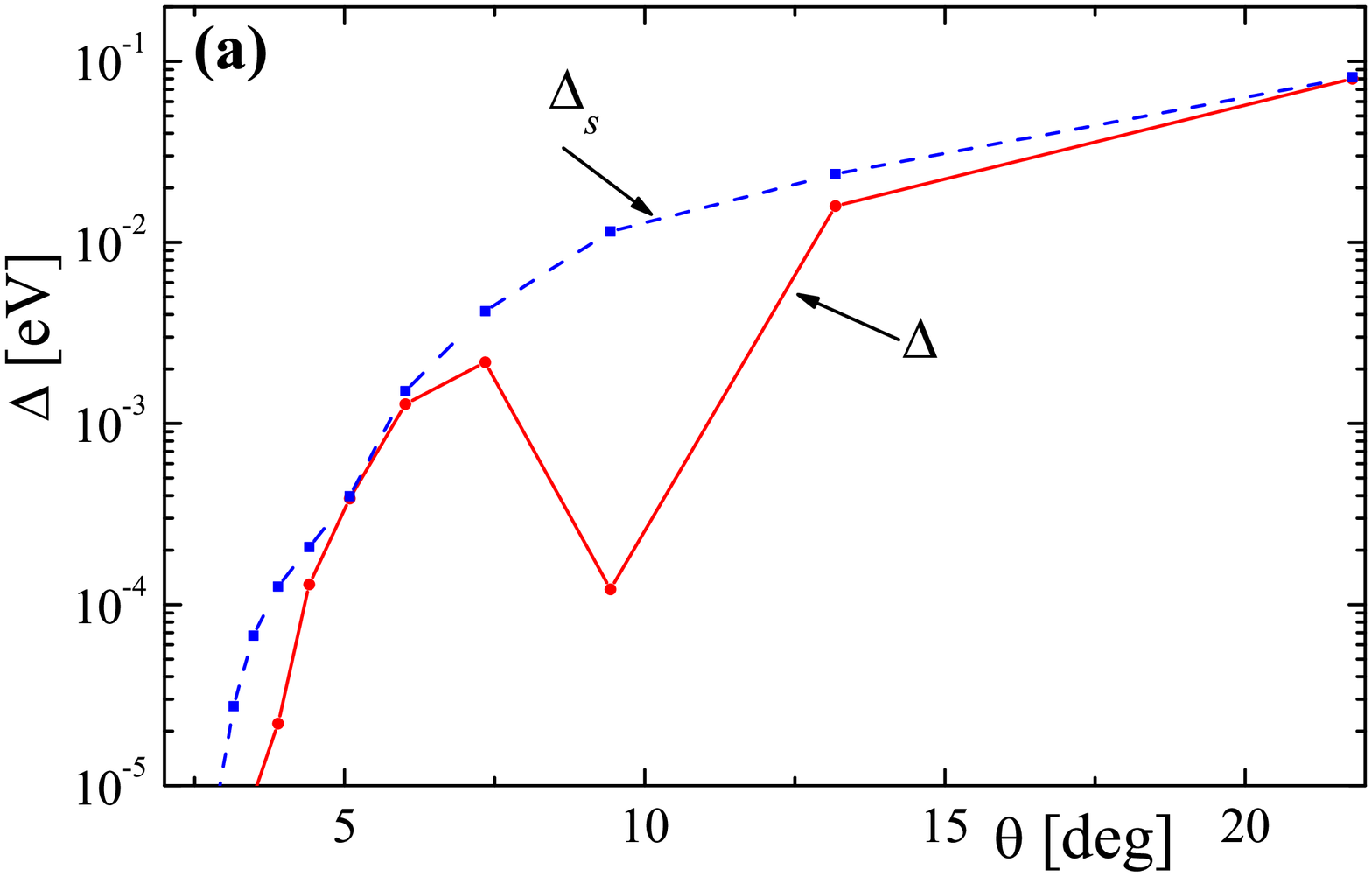}\hspace{2mm}\\
\includegraphics[width=1\columnwidth]{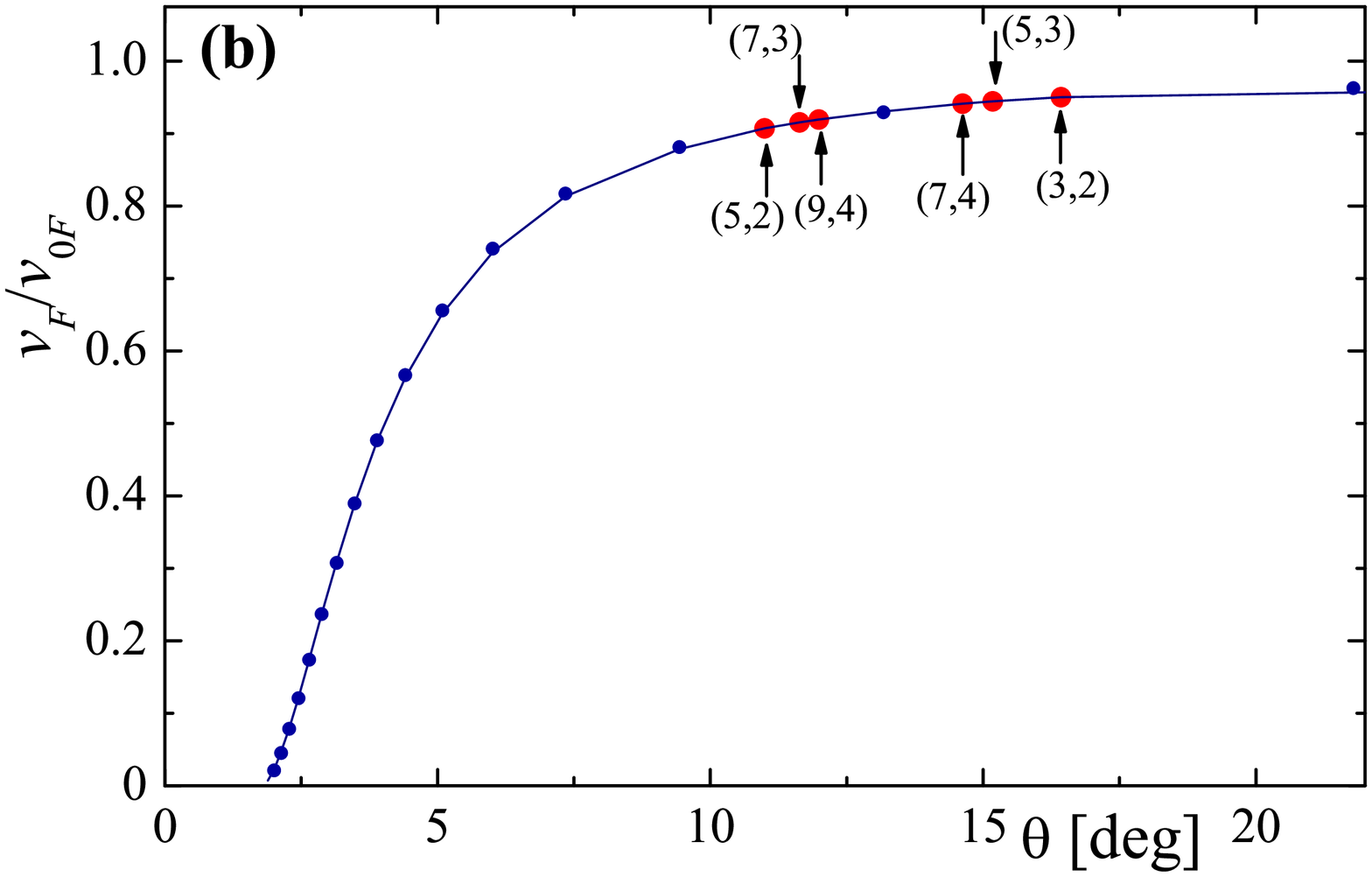}
\caption{(Color online) Single-electron spectrum properties as functions of
the twist angle $\theta$.
In panel~(a) the dependencies of the band gap $\Delta$ (red solid curve),
and band splitting
$\Delta_s$ (blue dashed curve) are shown for
$r=1$
structures.
In panel~(b) the Fermi velocity
$v_F$
for both
$r=1$
structures (small blue dots), and
$r\ne 1$
structures (larger red dots)
is plotted. To extract the gap $\Delta$, splitting
$\Delta_s$,
and Fermi velocity
$v_{\rm F}$,
the numerically determined low-energy bands
$\bar{E}_{\mathbf{k}}^{(\nu)}$
were fitted by
Eq.~\eqref{Efit}.
}
\label{FigDeltaV}
\end{figure}

\section{Large twist angles $\theta > \theta_c$}
\label{sect::large_angle}

\subsection{Superstructures with $r=1$}

We explained in
Sec.~\ref{Geometry}
that the superstructures with
$r>1$
can be viewed as almost-periodic repetitions of superstructures with
$r=1$.
Consequently, some electronic properties of a bilayer with generic values
of $\theta$ may be easily linked to the properties of a
$r=1$
system. This makes the study of the
$r=1$
case particularly useful. Below we calculate the spectra for several such
superstructures: we vary
$m_0$
from
$m_0=1$
($\theta=21.787^{\circ}$,
number of sites in the supercell
$N=28$)
to
$m_0=25$
($\theta=1.297^{\circ}$, $N=7804$).
The results for three different angles are shown in
Fig.~\ref{FigSpec}(a)--(c).
The results are
qualitatively different for $\theta$ larger and smaller than the critical
value $\theta_c\cong1.89^{\circ}$ corresponding to $m_0=17$.

For $r=1$ the number of low-energy bands $n_0=4$ for any $\theta$. When $\theta>\theta_c$ ($m_0<17$), two pairs of bands come close to the
Fermi level $\mu$ in the vicinity of the tBLG Dirac points
$\mathbf{K}_1$ and
$\mathbf{K}_2$; one pair from below and another
pair from above. The bands in each pair are almost degenerate in a large
range of momentum space. The smaller $\theta$, the smaller the energy
difference between bands in these pairs. Neither of the bands reach the
Fermi energy for $\theta>\theta_{\rm c}$. Thus, in this case the system is
an insulator with a non-zero band gap $\Delta$.
Near the Dirac points $\mathbf{K}_{1,2}$, at $\mathbf{k}=\mathbf{K}_{1,2}+\bm{\delta k}$,
the energy spectrum can be approximated as
\begin{equation}\label{Efit}
\bar{E}_{\mathbf{K}_{1,2}+\bm{\delta k}}^{(\nu)}=\mu\pm\sqrt{\Delta^2+v_F^2\left(|\bm{\delta k}|\pm k_0\right)^2}\,,
\end{equation}
where the different signs correspond to different bands; $\Delta$, $v_F$,
and $k_0$ are fitting parameters [see Fig.~\ref{FigSpec}(d, e)]. The
quantities $\Delta$ and
$v_F$,
calculated by fitting the numerical data for
$\bar{E}_{\mathbf{k}}^{(\nu)}$
using Eq.~\eqref{Efit},
are shown in
Fig.~\ref{FigDeltaV}
as functions of $\theta$.  The gap monotonously decreases when $\theta$
decreases, with a single exception at
$\theta\cong9.43^\circ$ ($m_0=3$). The gap $\Delta\gtrsim1$\,K, if $\theta\geqslant4.408^{\circ}$
($m_0\leqslant7$) and achieves the value $\Delta\cong0.08$\,eV when $\theta\cong21.787^{\circ}$ ($m_0=1$).
Thus, it can be experimentally measured if the twist angle is not small. In
the region
$\theta_c<\theta\lesssim4.4^{\circ}$,
the gap is too small, and one can consider the tBLG to be a semimetal. The
spectrum of the tBLG with a gap was observed in recent
experiments~\cite{ARPES_NatMat}.
However, the nature of this gap is unclear.

If we neglect the values $\Delta$ and $k_0$ in Eq.~\eqref{Efit}, the band
structure reduces to two doubly-degenerate Dirac cones located at the
points $\mathbf{K}_1$ and $\mathbf{K}_{2}$ and intersecting at higher
energies.
The Fermi velocity $v_F$ is smaller than that of the single-layer graphene,
$v_{0F}=ta\sqrt{3}/2$,
and it monotonously decreases with decreasing twist angle (see
Fig.~\ref{FigDeltaV}).
This picture is quite consistent with many previous studies utilizing
different
approaches~\cite{Pankratov1,Pankratov2,Pankratov3,Pankratov4,NanoLett,Morell,MeleReview,dSPRL,dSPRB,PNAS}.

The gap may be viewed as a consequence of hybridization between the states
located at the Dirac points of two graphene layers
$\mathbf{K}_{\theta}$
and
$\mathbf{K}'$
(and
$\mathbf{K}$
and
$\mathbf{K}'_{\theta}$).
Indeed, as it was mentioned above, for commensurate structures with
$r\neq3n$
the momenta
$\mathbf{K}'$
and
$\mathbf{K}_{\theta}$
are equivalent to each other [see Eq.~\eqref{Kn3n}].
The matrix element mixing the states near
$\mathbf{K}_{\theta}$
and
$\mathbf{K}'$
is allowed by symmetry, which leads to the band splitting and gap opening.

This hybridization is ignored in the continuum
approximations~\cite{dSPRL,dSPRB,PNAS},
even though Refs.~\onlinecite{dSPRL,dSPRB}
mentioned such a possibility. The phenomenological approach taking into
account the hybridization between different Dirac cones in the tBLG was
proposed in
Ref.~\onlinecite{MelePRB2010}.
However, if
$r=1$,
as we assume in this subsection, the formalism of
Ref.~\onlinecite{MelePRB2010}
predicts a gapless spectrum, failing to capture the insulating state of
$r=1$
structures.

A more general result of the hybridization between electron states near the
Dirac points is the breakdown of the double degeneracy of the low-energy
bands of the tBLG. For $r=1$ structures, the band gap $\Delta$ would be the
measure of such a band splitting. However, this is not so for structures
with $r=3n$, where the gap is zero as we will show below. Here, following
Refs.~\onlinecite{Pankratov2,Pankratov4} we introduce the quantity
\begin{equation}
\Delta_{s}=(\bar{E}_{\mathbf{K}}^{(4)}-\bar{E}_{\mathbf{K}}^{(1)})/2
\end{equation}
as a measure of this band splitting. According to Eq.~\eqref{Efit}, $\Delta$ and
$\Delta_s$ satisfy the relationship
$$
\Delta_s=\sqrt{\Delta^2+v_F^2k_0^2}\,.
$$
The dependence of $\Delta_s$ on $\theta$ is also shown in
Fig.~\ref{FigDeltaV}. In contrast to the band gap, $\Delta_s$ monotonously decreases
with the twist angle $\theta$.

\begin{figure}
\includegraphics[width=0.9\columnwidth]{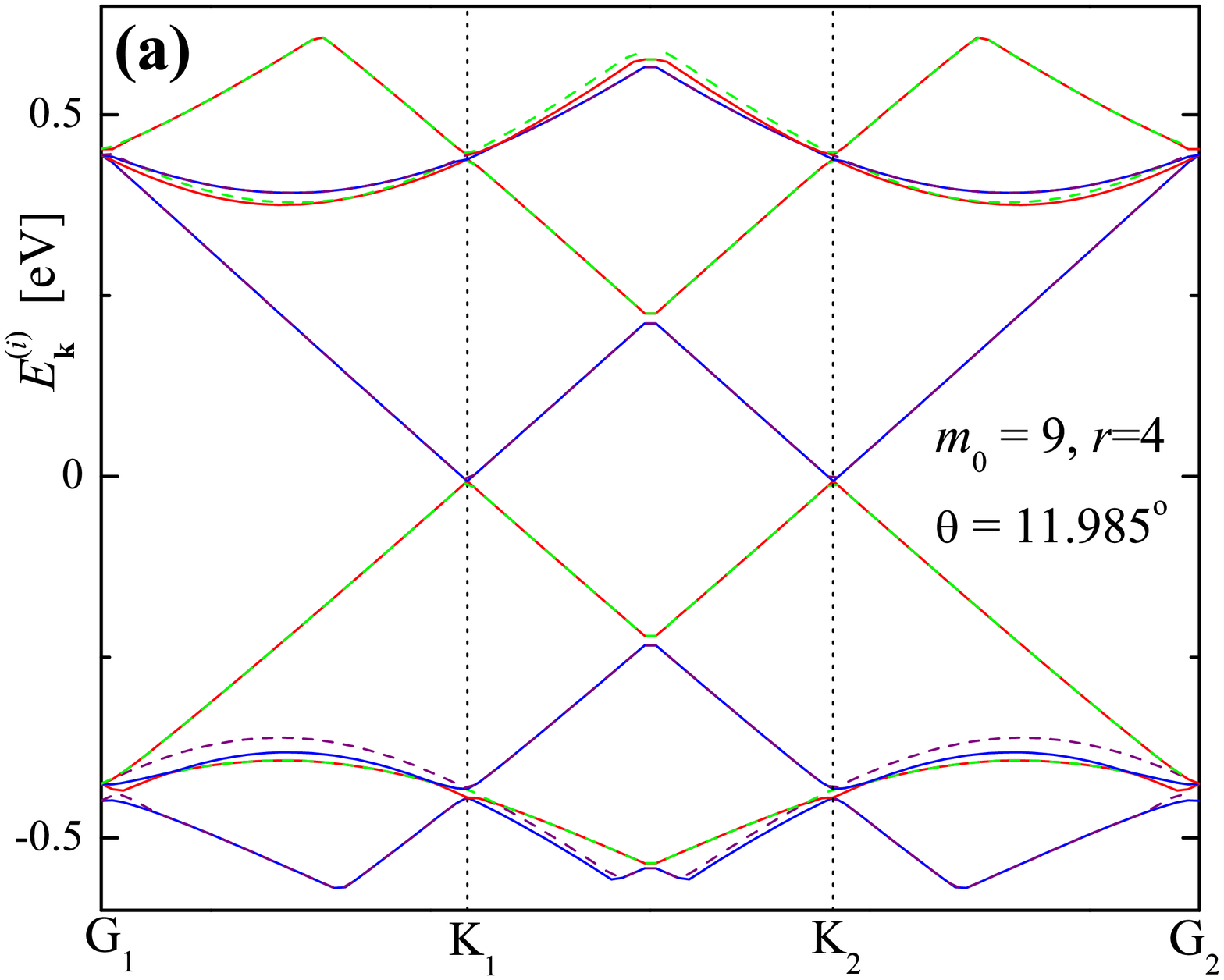}\vspace{2mm}\\
\includegraphics[width=0.9\columnwidth]{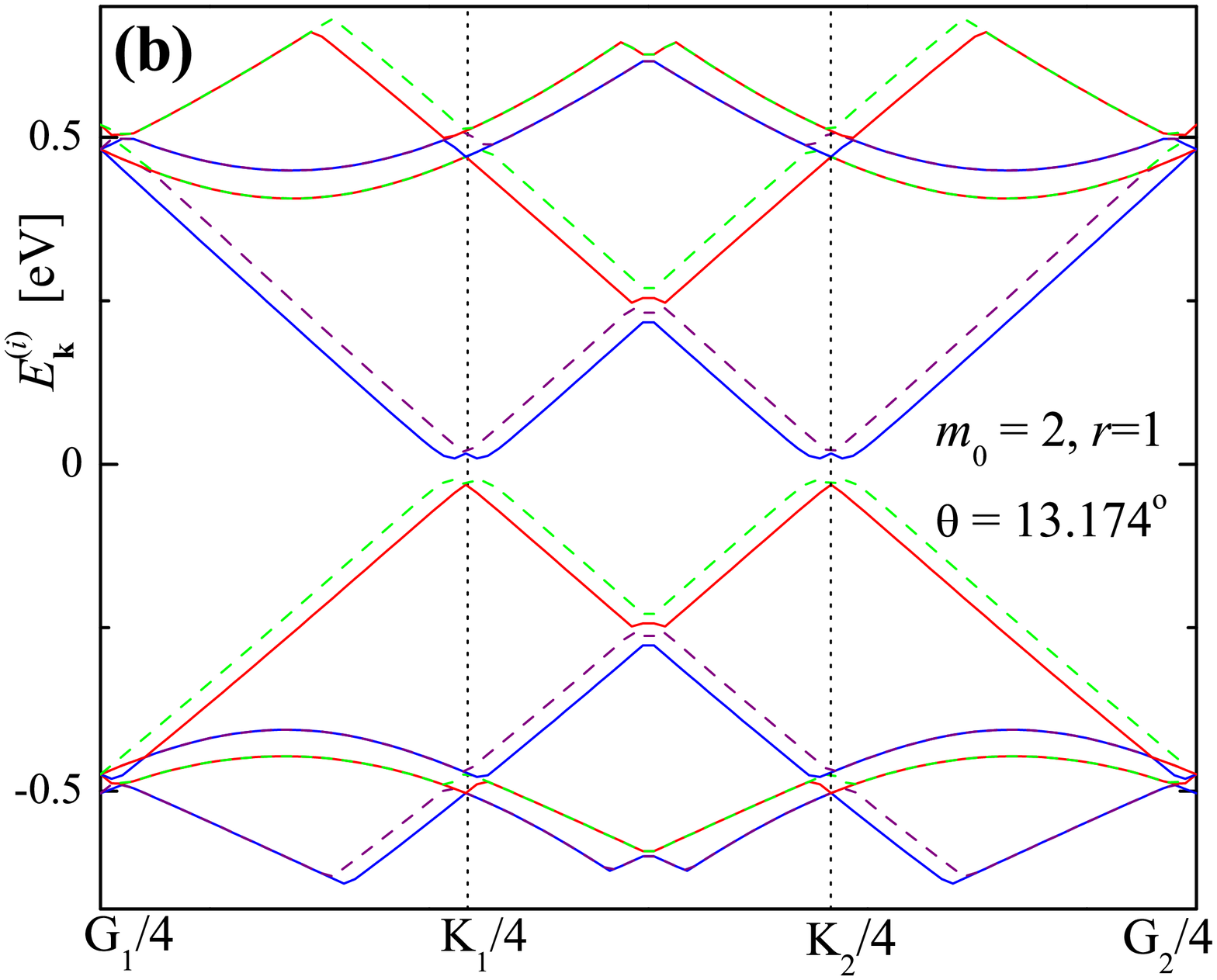}
\caption{(Color online) The spectra of the twisted bilayer graphene
calculated for structures
$(9,4)$
[panel (a)]
and
$(2,1)$
[panel (b)]. The spectrum for
$(9,4)$
is calculated along the line connecting the points
$\mathbf{G}_1(9,4)$
and
$\mathbf{G}_2(9,4)$.
The spectrum for
$(2,1)$
is calculated in the folded (four times) reciprocal cell of the superlattice
along the line connecting the points
$\mathbf{G}_1(2,1)/4\approx\mathbf{G}_1(9,4)$
and
$\mathbf{G}_2(2,1)/4\approx\mathbf{G}_2(9,4)$.}
\label{FigSpecRneq1}
\end{figure}

The tight-binding calculations in
Refs.~\onlinecite{Pankratov2,Pankratov4}
predicted the existence of the band splitting in the tBLG. However, the
value of
$\Delta_s$
is at least one order of magnitude smaller than our value (c.f.,
Fig.~\ref{FigDeltaV}
with
Fig.~9 in Ref.~\onlinecite{Pankratov2}, or with Fig.~1 in
Ref.~\onlinecite{Pankratov4}). We attribute this discrepancy to the
different choice of the function $t_\bot (\mathbf{r};\mathbf{r'})$. We
believe that our choice is more suitable for the tBLG since it reproduces the
SWM hopping parameters for AB bilayer graphene. A similar conclusion was
reached in Ref.~\onlinecite{MelePRB2011}. The value $\Delta_{s}$ estimates
the band splitting near the Dirac point. However, it can be substantially
larger in other regions of momentum space. The band splitting was
experimentally observed by ARPES measurements in
Ref.~\onlinecite{ARPES_PRB}.

\subsection{Superstructures with $r\neq1$}
\label{SectionRneq1}


The superstructures with
$r=1$,
considered in the previous subsection, exhaust a fairly limited set of twist angles. Can the knowledge about this set be sufficient to adequately capture
the properties of the tBLG for a generic value of
$\theta$?
The answer to this question is positive, if one aims to describe the Fermi
velocity (see, for example,
Fig.~\ref{FigDeltaV}).
However, as we will see below, it is negative, if one needs to know
the band gap. Therefore, a detailed study of
$r\ne 1$
systems is required.

The supercell of the structure $(m_r,r)$ with $r>1$
contains $r^2/g$ Moir\'{e} cells, where $g=1$ if $r\neq3n$, or
$g=3$
otherwise. The arrangements of carbon atoms inside these cells are
slightly different from each other and approximately correspond to the
$r=1$ superstructure with $m_0=[m_r/r]$, where $[a]$ means the integer part of
$a$.
Since these structures are not completely identical, this can affect
the electronic structure of the tBLG. In this subsection we consider the
differences and similarities between electronic spectra of `basic'
$r=1$
structures and superstructures with $r>1$.

For the structure
$(m_r,r)$
and close `basic' structure
$(m_0,1)$
with
$m_0=[m_r/r]$,
we have
\begin{equation}
\mathbf{G}_{1,2}(m_r,r)\approx\mathbf{G}_{1,2}(m_0,1)/r
\end{equation}
if $r\neq3n$, or
\begin{eqnarray}
\mathbf{G}_{1}(m_r,r)&\approx&\left[\mathbf{G}_{1}(m_0,1)-\mathbf{G}_{2}(m_0,1)\right]/r\,,\nonumber\\
\mathbf{G}_{2}(m_r,r)&\approx&\left[\mathbf{G}_{1}(m_0,1)+2\mathbf{G}_{2}(m_0,1)\right]/r
\end{eqnarray}
otherwise. For $r\neq3n$ we can compare the low-energy spectra of the
$(m_r,r)$
and
$(m_0,1)$
structures directly by folding the Brillouin zone of the
$(m_0,1)$
superstructure.
Figure~\ref{FigSpecRneq1}
shows the low-energy spectra of the structure
$(9,4)$
with
$\theta=11.985^{\circ}$
and the structure $(2,1)$ having a similar twist angle $\theta=13.174^{\circ}$.
The spectrum for $(9,4)$ is calculated along the line connecting the
reciprocal supercell vectors of this structure:
$\mathbf{G}_{1}(9,4)$
and
$\mathbf{G}_{2}(9,4)$.
The spectrum for
$(2,1)$
is calculated along
the line connecting the points
$\mathbf{G}_{1}(2,1)/4$
and
$\mathbf{G}_{2}(2,1)/4$ in the reciprocal cell of the superlattice, which has been folded four times.
After folding, the momenta
$\mathbf{k}+i\mathbf{G}_1/4+j\mathbf{G}_2/4$ (with $i,j=0,1,2,3$)
become equal to each other, and the number of bands increases $16$ times. We see that these
spectra are very similar to each other, with the single exception that the
splitting of the low-energy bands for the
$(9,4)$ structure is much smaller than that for the $(2,1)$
structure.

\begin{figure}[t]
\includegraphics[width=0.8\columnwidth]{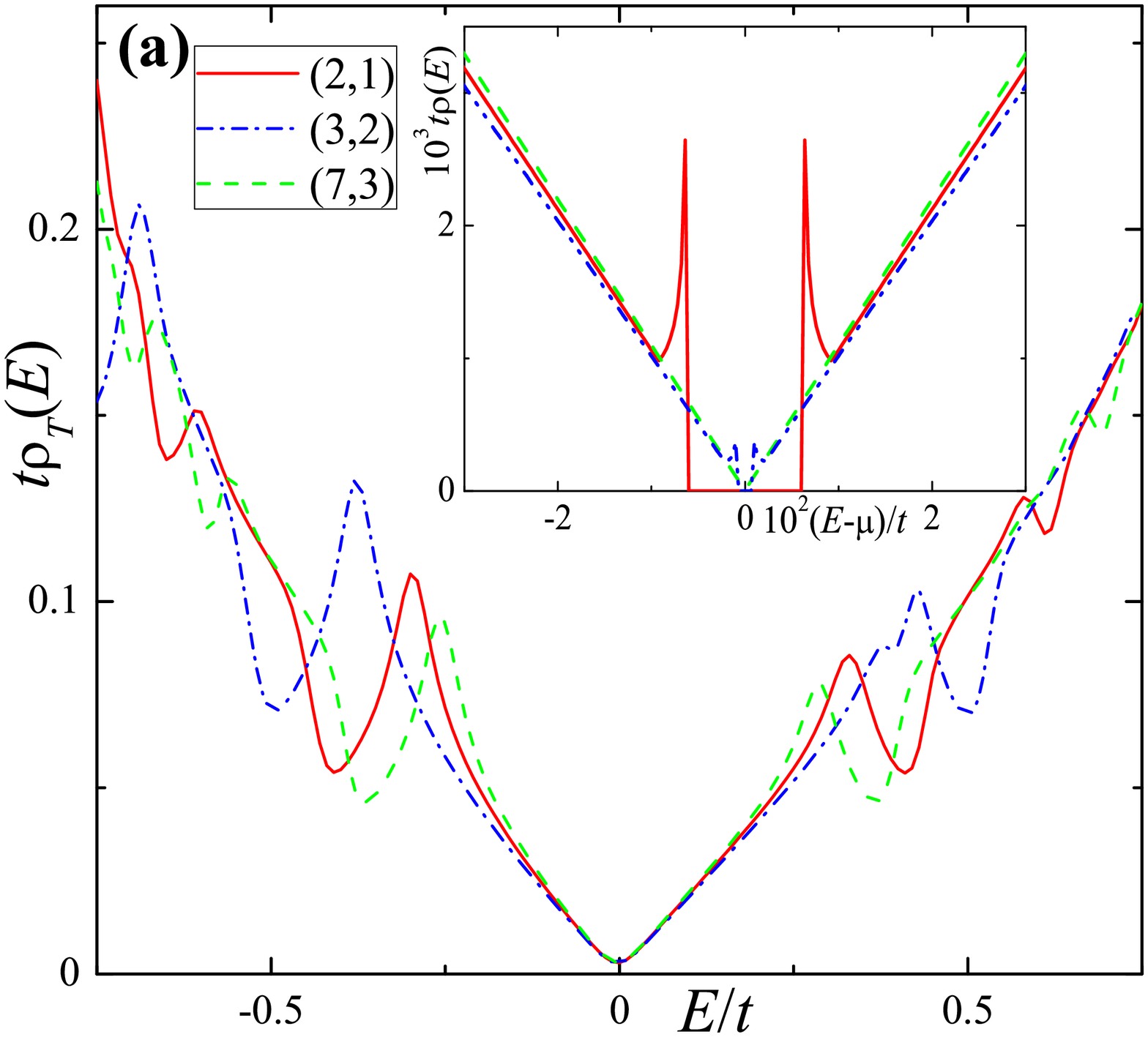}\\
\includegraphics[width=0.83\columnwidth]{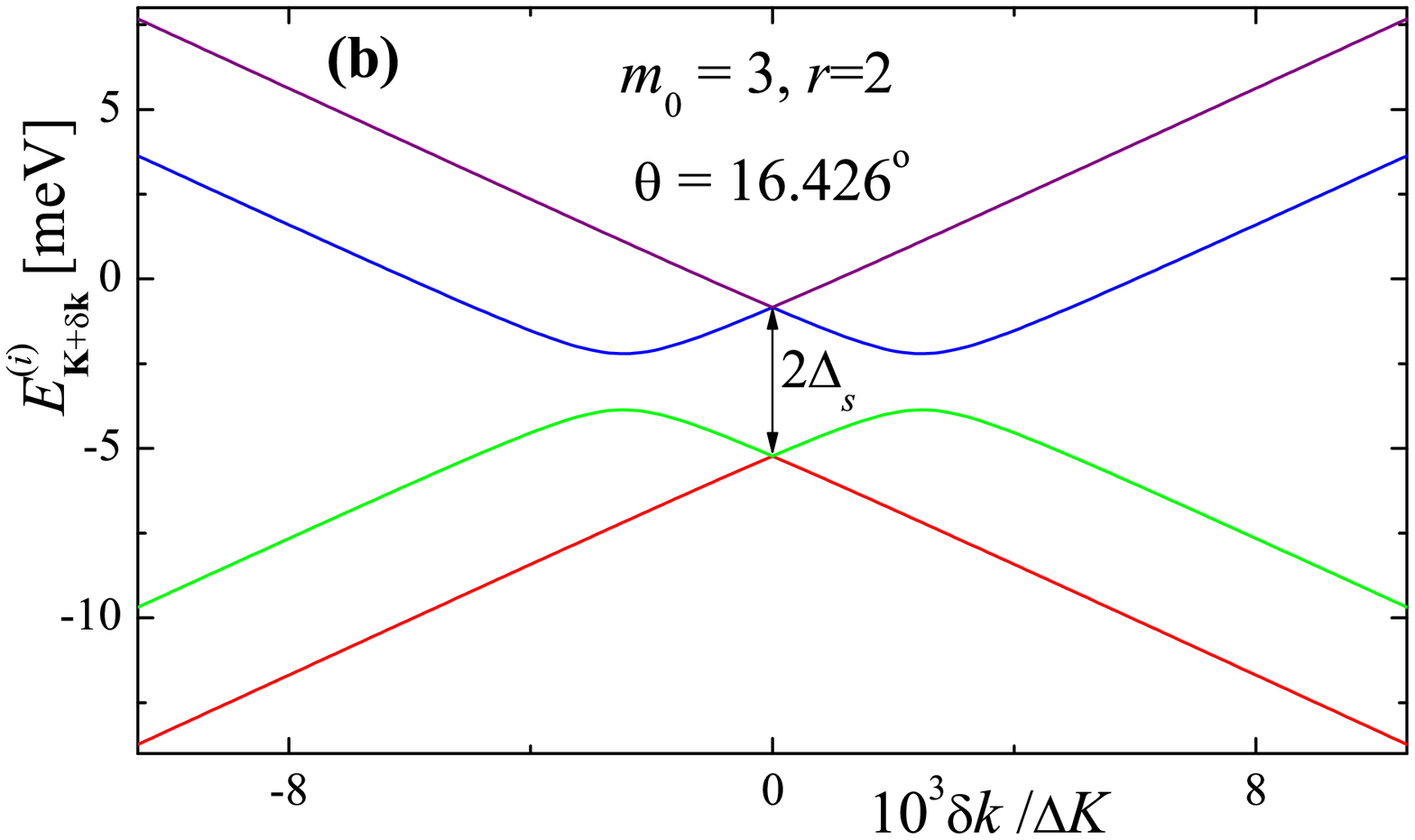}\\
\includegraphics[width=0.82\columnwidth]{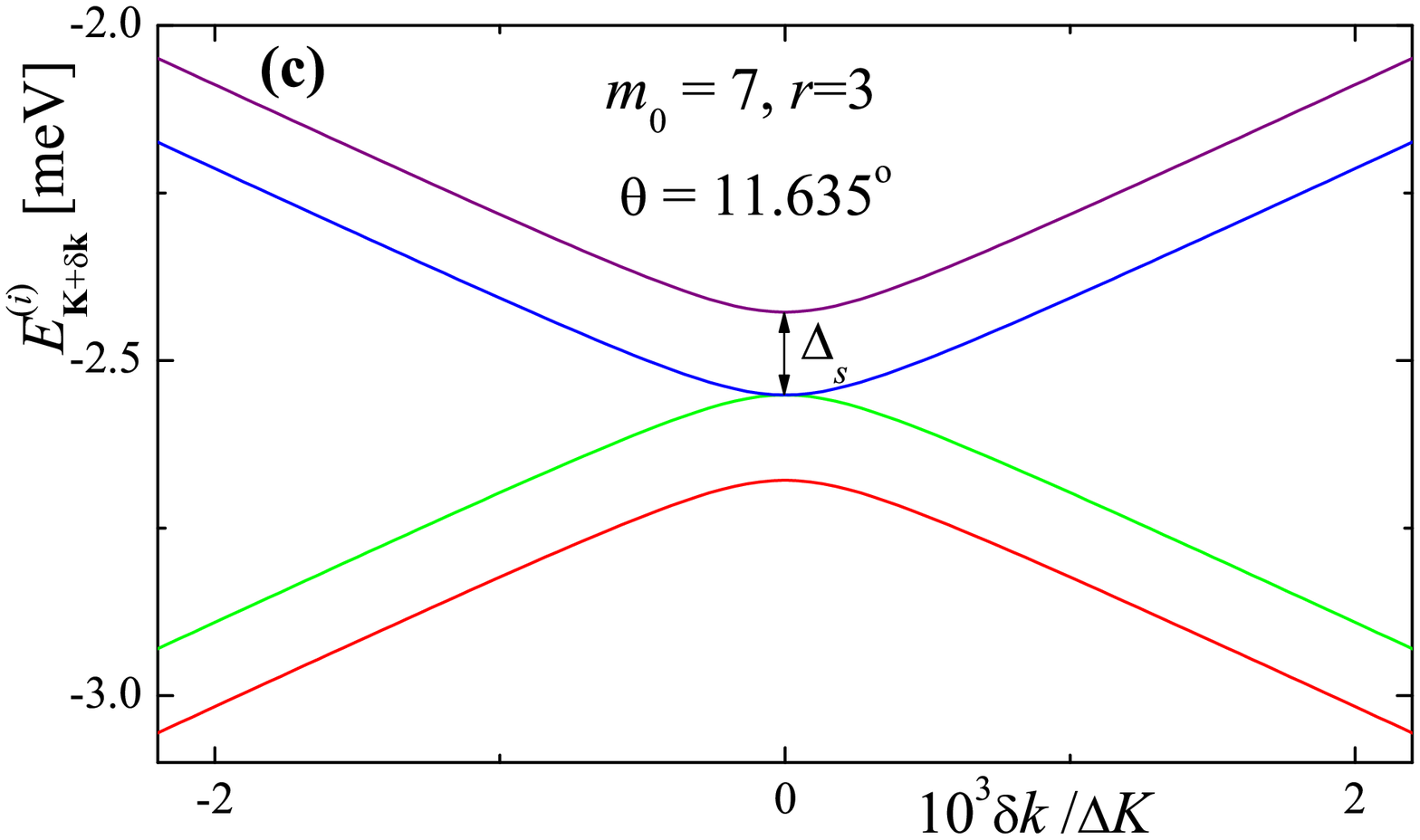}
\caption{(Color online) (a) Finite temperature ($T=0.01t$) density of
states $\rho_T(E)$ calculated for the superstructure
$(2,1)$
($\theta\approx13.174^{\circ}$),
the superstructure
$(3,2)$
($\theta\approx16.426^{\circ}$),
and the superstructure
$(7,3)$
($\theta\approx11.635^{\circ}$).
The peaks in the range
$0.25< |E/t| < 0.5$
are van Hove singularities due to the overlapping of two tBLG Dirac
cones. The inset shows the zero-temperature density
of states $\rho$ calculated for energies close to the Fermi level. The structures
$(2,1)$ and $(3,2)$ are gapped; the gap for the structure $(3,2)$ is much lower
than that for the $(2,1)$.  The structure $(7,3)$  is gapless and has a finite
density of states at the Fermi level. (b)\,--\,(c) Typical tBLG spectra
close to the Dirac point for $r\neq3n$ (b) and $r=3n$ (c) structures. The
value of the band splitting parameter $\Delta_s$ is shown by double
arrows.}\label{FigDOS}
\end{figure}

The direct comparison of the spectra for $(m_r,r)$ and $(m_0,1)$ structures
with $r=3n$ in the way described above is not possible. This is because for
$r=3n$, the folding procedure brings the Dirac points of the
$(m_0,1)$
superlattice, $\mathbf{K}_{1,2}$, to the single $\bm{\Gamma}(0,0)$ point.
For $(m_r,r)$ structures, however, the Dirac points
$\mathbf{K}_{1,2}$
are not equivalent to each other, their locations are given by
Eq.~\eqref{K12}.
In principle, one can fold both structures in such a manner that the
corresponding folded Brillouin zones coincide. However, this increases
drastically the number of low-energy bands to compare. Instead, we compare
the density of states (DOS) for different superstructures with close twist
angles. More precisely, using numerical integration over momentum
space, we calculate the density of states at finite temperature $T$:
\begin{equation}\label{rhoT}
\rho_T(E)=
\frac{1}{N}
\sum_{i}\int\!\!
\frac{d^2\mathbf{k}}{\upsilon_{\text{BZ}}}\,
\frac{1}{4T\cosh^2\!\!\left(\frac{E-E^{(i)}_{\mathbf{k}}}{2T}\right)}\,.
\end{equation}
The density of states is normalized such that $\int_{-\infty}^{+\infty}dE\,\rho_T(E)=1$. Typical curves (for $\theta>\theta_c$) are presented
in Fig.~\ref{FigDOS} (we choose $T=0.01t$). The DOS behaves almost linearly at small energies and
have several van Hove peaks at larger energies. The two van Hove peaks
closest to zero energy are due to the overlapping of two tBLG Dirac cones,
which leads to the appearance of saddle points in the spectrum. Such a
behavior of the DOS is in agreement both with theoretical
studies~\cite{dSPRL,dSPRB} and STM measurements.~\cite{STM_VHS,STM_VHS2}
Comparing the curves $\rho_T(E)$ for structures with close twist angles
(see Fig.~\ref{FigDOS}), we find that the density of states continuously
depends on $\theta$ when $T>\Delta_s$. This is not so, however, at smaller
temperatures and energies very close to the chemical potential $\mu$ at
zero doping. To determine the density of states in this case we have to
calculate the spectrum near the Dirac points. For $r\neq3n$ structures, the
low-energy bands are described with very good accuracy by
Eq.~\eqref{Efit}, while for $r=3n$ the momentum dependence of the
low-energy bands is [see Fig.~\ref{FigDOS}(b,c)]:
\begin{eqnarray}\label{EfitReq3}
\bar{E}_{\mathbf{K}+\bm{\delta k}}^{(1,4)}&=&\mu\mp\sqrt{\Delta_s^2+v_F^2\bm{\delta k}^2}\,,\nonumber\\
\bar{E}_{\mathbf{K}+\bm{\delta k}}^{(2,3)}&=&\mu\mp\left(\sqrt{\Delta_s^2+v_F^2\bm{\delta k}^2}-\Delta_s\right)\,.
\end{eqnarray}

\begin{figure}[t]
\includegraphics[width=0.95\columnwidth]{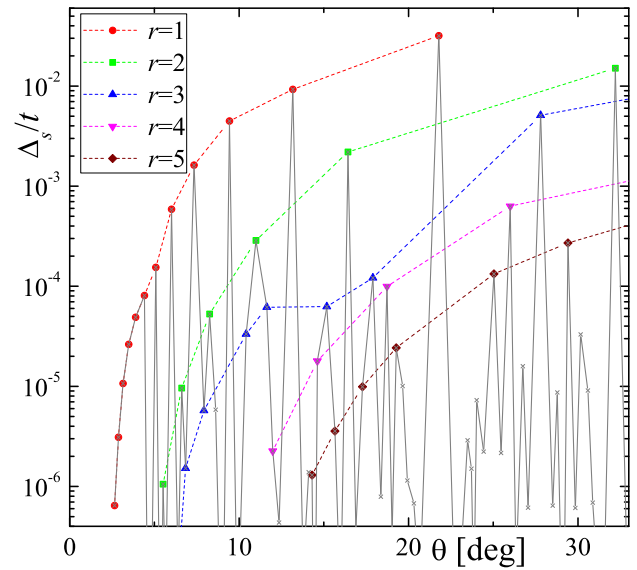}
\caption{(Color online) The band splitting
$\Delta_s$
as a function of the twist angle $\theta$ calculated for all
superstructures with
$N<2000$
(gray solid line). Different dashed curves connect the points corresponding
to superstructures with the same value of $r$. We can see that, if $r$ is
fixed, the band splitting decreases monotonously as
$m_0$
grows (when
$m_0$
grows, the angle $\theta$ decreases). However, when $r$ is not restricted, the
splitting
$\Delta_s$
can change exponentially for weak variations of $\theta$.
}
\label{FigSplitting}
\end{figure}

Thus, for
$r\neq3n$
structures, the spectrum has a gap $\Delta$, while for
$r=3n$
the spectrum is gapless, and the bands
$\bar{E}_{\mathbf{K}+\bm{\delta k}}^{(2,3)}$
touch each other at the Dirac points. For these structures, the density of
states is non-zero at the Fermi level and is proportional to the
band splitting $\Delta_s$. For energies close to the Fermi level, the
density of states (at $T=0$) for these two types of structures can be
written as
\begin{eqnarray}
\rho_0(E+\mu)&=&\frac{a^2\sqrt{3}|E|}{8\pi v_F^2}\left[\Theta(|E|-\Delta_s)+\phantom{\frac{k_0}{\sqrt{E^2}}}\right.\\
&&\left.\frac{v_Fk_0}{\sqrt{E^2-\Delta^2}}\Theta(|E|-\Delta)\Theta(\Delta_s-|E|)\right]\,,\nonumber
\end{eqnarray}
if $r\neq3n$, or
\begin{equation}
\rho_0(E+\mu)=\frac{a^2\sqrt{3}}{16\pi v_F^2}\Big[|E|+\Delta_s+|E|\Theta(|E|-\Delta_s)\Big]\,,
\end{equation}
if $r=3n$.

Such a behavior of the low-energy bands for $r\neq3n$ and $r=3n$ structures
coincides with tight-binding calculations done in
Ref.~\onlinecite{Pankratov4}.
However, the low-energy approach proposed in
Ref.~\onlinecite{MelePRB2010}
gives the opposite results: the spectrum of the
$r=3n$
system is described by
Eqs.~\eqref{Efit},
while for
$r\neq3n$
the band structure corresponds to
Eq.~\eqref{EfitReq3}.

The DOS at zero temperature near the Fermi level for several
superstructures with nearby twist angles is shown in the inset in
Fig.~\ref{FigDOS}(a).
At small energies, the density of states exhibits strong sensitivity to
the type of structure, while for
$|E-\mu|>\max(\Delta_s)$, the DOS for similar twist angles
almost coincide with each other.

In the energy region, where the DOS curves coalesce, the density of states
depends linearly on the energy. The proportionality coefficient is set by
the Fermi velocity, which is determined by fitting the numerically
calculated spectrum with either
Eq.~\eqref{Efit} or~\eqref{EfitReq3}
(the choice between these two equations is based on the structure type).
For several
$r\neq1$
structures, whose twist angles are close to the twist angle $\theta$ of the
$(2,1)$
structure, the Fermi velocity is shown as circles in
Fig.~\ref{FigDeltaV}. We see that the Fermi velocities of the
$r=1$
and
$r\neq1$
structures are well described by a single smooth curve monotonously
decreasing with decreasing twist angle.

\begin{figure}[t]
\includegraphics[width=0.9\columnwidth]{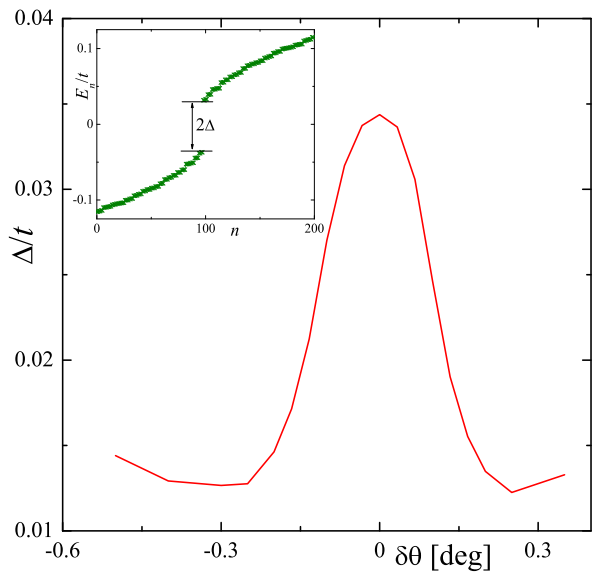}
\caption{(Color online) Dependence of the band gap $\Delta$ on the deviation $\delta\theta$ of the twist angle from the value $\theta(1,1)\cong21.787^{\circ}$, calculated for finite sample of rhombic shape containing $151\times151$ unit cells in each graphene layer. The total number of atoms in the sample $N_{\text{atoms}}=91,\!204$. The inset shows the energy distribution of the first $200$ electron levels close to the zero energy, calculated for $\delta\theta=0$.}\label{FigGapFin}
\end{figure}

Thus, our analysis indicates that the density of states at finite
temperatures (for $T>\Delta_s$) and the Fermi velocity can be considered as
continuous functions of the twist angle. The band splitting, and even the
type of $T=0$ low-energy spectrum, however, are very sensitive to the type of
superstructure, and can vary significantly for structures with arbitrary
close twist angles. Let us discuss this issue in more details. In
Fig.~\ref{FigSplitting}
we plot the band splitting
$\Delta_s$
as a function of the twist angle $\theta$ for all superstructures, whose
supercell contains
$N<2000$
atoms. It is seen from this figure, that
$\Delta_s$
is not a monotonous function of $\theta$: any small deviation of the twist
angle from a given value changes drastically the band splitting. However,
there is some order in this chaos: the band splitting
$\Delta_s$
for superstructures with fixed $r$ monotonously decreases when the twist
angle decreases. All curves
$\Delta_s$
versus $\theta$ at fixed $r$ are qualitatively similar to each other, and
the curve for
$r=1$
superstructures lies above all other curves.

Thus, among all possible superstructures in some range of twist angles, the
maximum band splitting corresponds to the $r=1$ superstructure. According to
our calculations, for $m_0<7$ the band splitting exceeds $1$ Kelvin, which
is experimentally observable.

However, the discontinuous behavior of
$\Delta_s$
versus $\theta$ makes the direct interpretation of the graph in
Fig.~\ref{FigSplitting}
problematic. After all, in any realistic situation the twist angle is known with
finite error. Examining
Fig.~\ref{FigSplitting}
we discover that within a given small interval of $\theta$ one can find an
exponentially wide range of band splittings.

To resolve this paradox one must remember that the data in
Fig.~\ref{FigSplitting}
is valid only for infinite and ideally clean samples with infinite
mean-free path. In an experimental situation these assumptions are not valid. Let
us denote a length scale $\tilde{L}$ characterizing the coherent motion of electrons in
the tBLG. This scale can be limited by the sample size for mesoscopic samples, or
the mean-free path of electrons scattered by impurities, phonons, etc.

Imagine now that the twist angle deviates from the value
$\theta_0$
corresponding to the
$(m_0,1)$
superstructure by some small quantity
$\delta\theta$.
Using Eqs.~\eqref{theta}--\eqref{Nsc}, one can show that the minimal supercell size
$L_{\text{sc}}$ among all $r\neq1$ superstructures inside this angular interval
can be estimated as:
\begin{equation}
\frac{L_{\text{sc}}}{a}\sim\frac{\theta_0}{|\delta\theta|}\,.
\end{equation}
Our calculations of the spectrum and the band
splitting are relevant only when $\tilde{L}\gg L_{\text{sc}}$. Thus, the results
for $\Delta_s$ presented here are not applicable for
\begin{equation}
|\delta\theta|\lesssim\theta_0\frac{a}{\tilde{L}}\,.
\end{equation}
Inside this region of twist angles we should take into account the influence of the electron scattering or the sample size on the band gap and band splitting. We expect, that this will lead to a continuous dependence of $\Delta$ on $\theta$.

To verify this we diagonalized the Hamiltonian~\eqref{H}
for finite-size samples. The sample has the shape of a rhombus with sides
$\tilde{L}$ and an acute angle between sides equal to $60^{\circ}$. The
rotation of the layer $2$ is performed around the central point of the
sample, where the $A1$ and $B2$ carbon atoms are located. We now choose
$(m_0,r)=(1,1)$,
which corresponds to
$\theta_0\cong21.787^{\circ}$,
as a reference structure. The sample size should be large enough in order
to suppress the size-quantization effect. At small energies, the energy
difference between neighboring electron states is about
$v_F/\tilde{L}$.
Thus, the parameter
$\tilde L$
must satisfy the inequality
$$\tilde{L}/a\gg v_F/(a\Delta)\sim t/\Delta\approx30
$$
for
$t=2.57$\,eV,
and the gap
$\Delta\cong0.08$\,eV
corresponding to the $(1,1)$ structure.

In addition to the size-quantization effect, we must deal with another
complication: the emergence of  low-energy states localized at the sample
edges. To eliminate them an extra term is added to the Hamiltonian: we
introduce the potential energy difference between $A$ and $B$ atoms which
decays exponentially fast away from the edge. This ``boundary condition"
pushes the edge states eigenenergies out of the low-energy range.

\begin{figure}[t]\centering
\includegraphics[width=0.95\columnwidth]{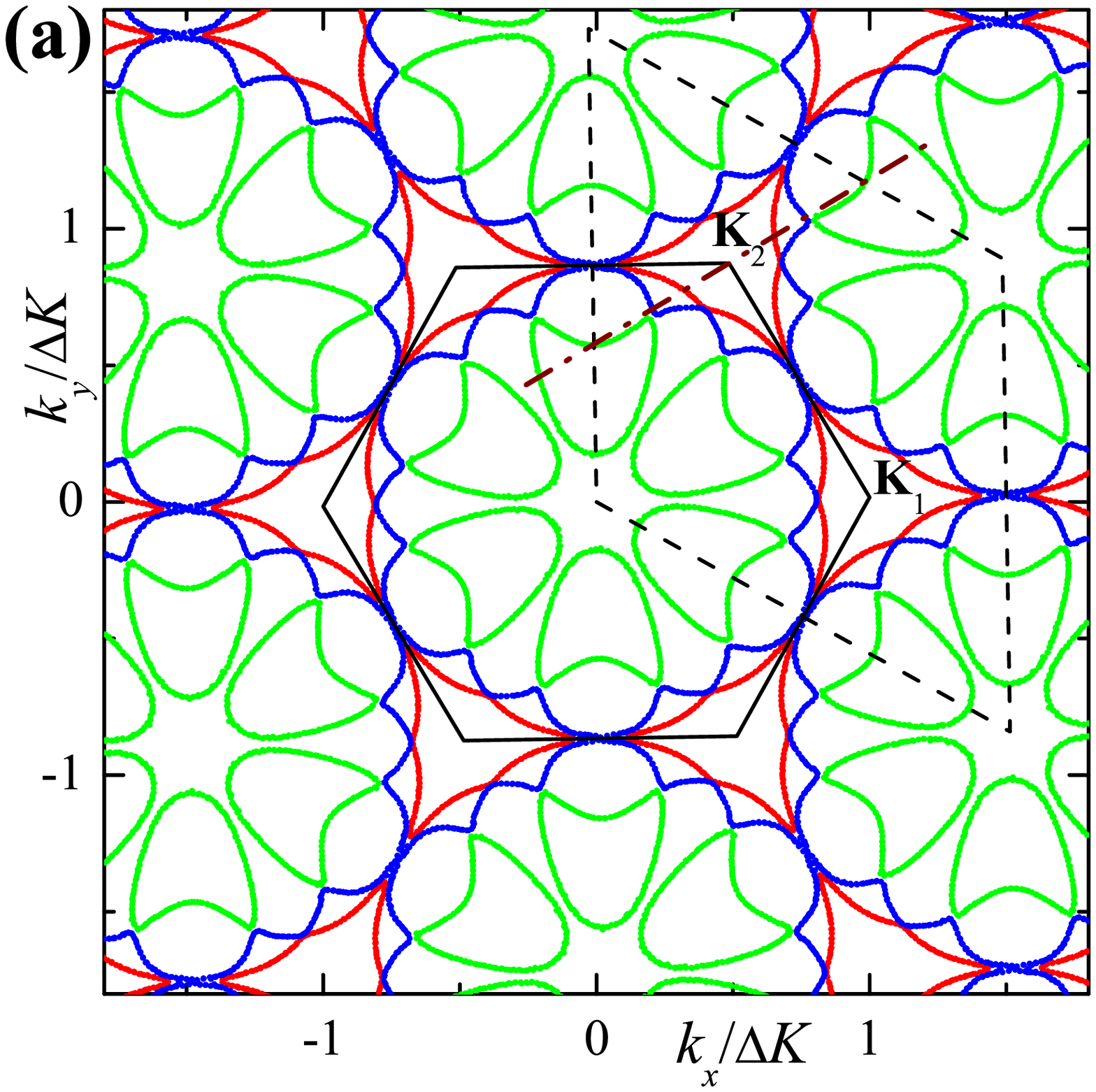}\\
\includegraphics[width=0.95\columnwidth]{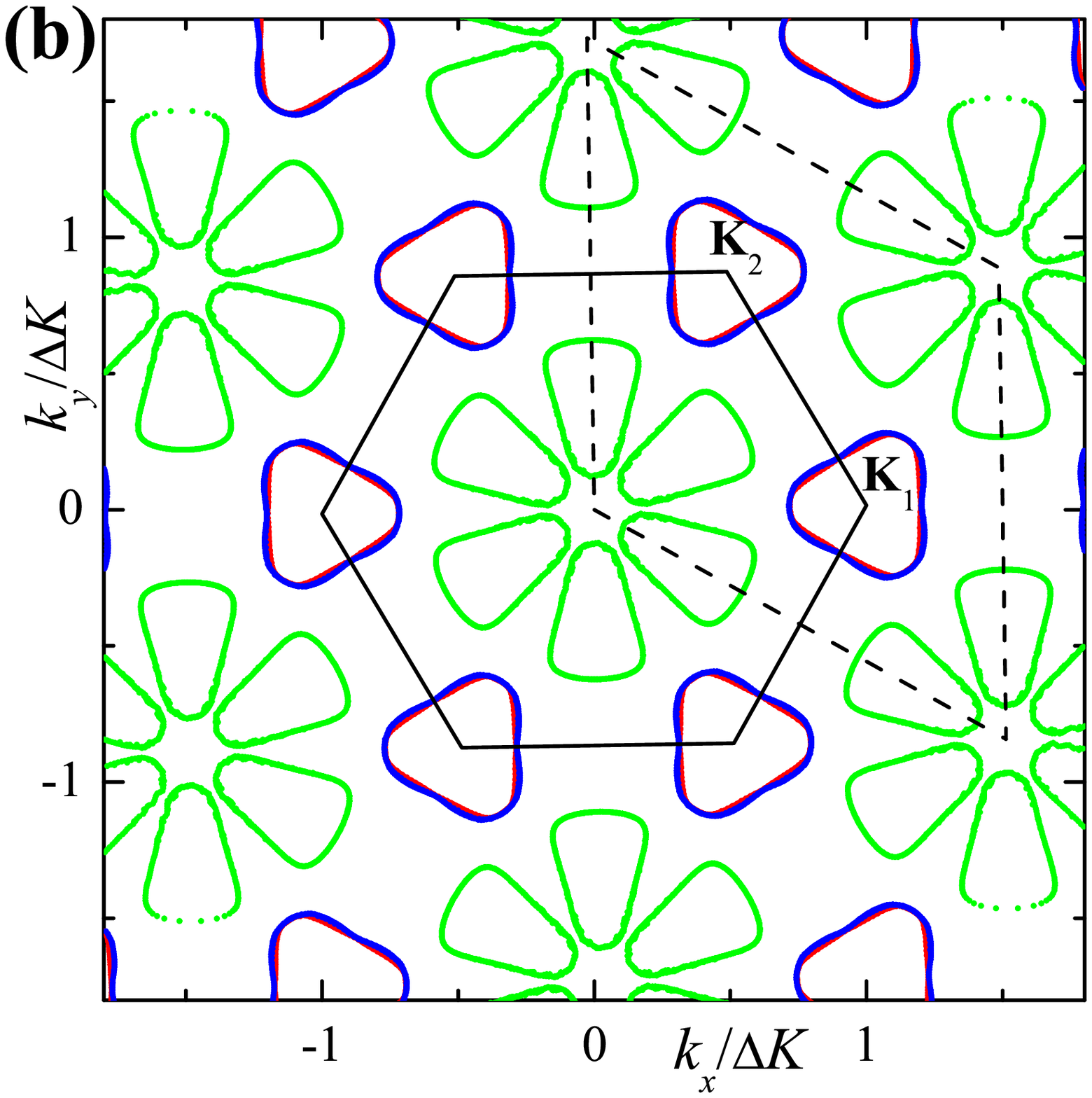}
\caption{(Color online) Fermi surfaces of the superstructures $(17,1)$
[$\theta\cong1.89^{\circ}$, panel (a)] and $(18,1)$
[$\theta\cong1.79^{\circ}$, panel (b)] calculated at half-filling.
Different colors correspond to different bands intersecting the Fermi
level $\mu$ at half-filling. The first Brillouin zone (hexagon) and the reciprocal supercell (rhombus) are
also shown. The dot-dashed line in panel (a) shows the way along which the
spectrum presented in
Fig.~\ref{FigSpec}(f)
is calculated.
}\label{FigFSr1}
\end{figure}

The band gap of the finite-size sample as a function of the deviation
$\delta\theta$
from the twist angle
$\theta_0$
is shown in
Fig.~\ref{FigGapFin}.
We see that $\Delta$ decreases continuously from its maximum at
$\delta\theta=0$
down to the background non-zero value set by the size quantization.

\begin{figure*}%
\centering
\includegraphics[width=0.46\textwidth]{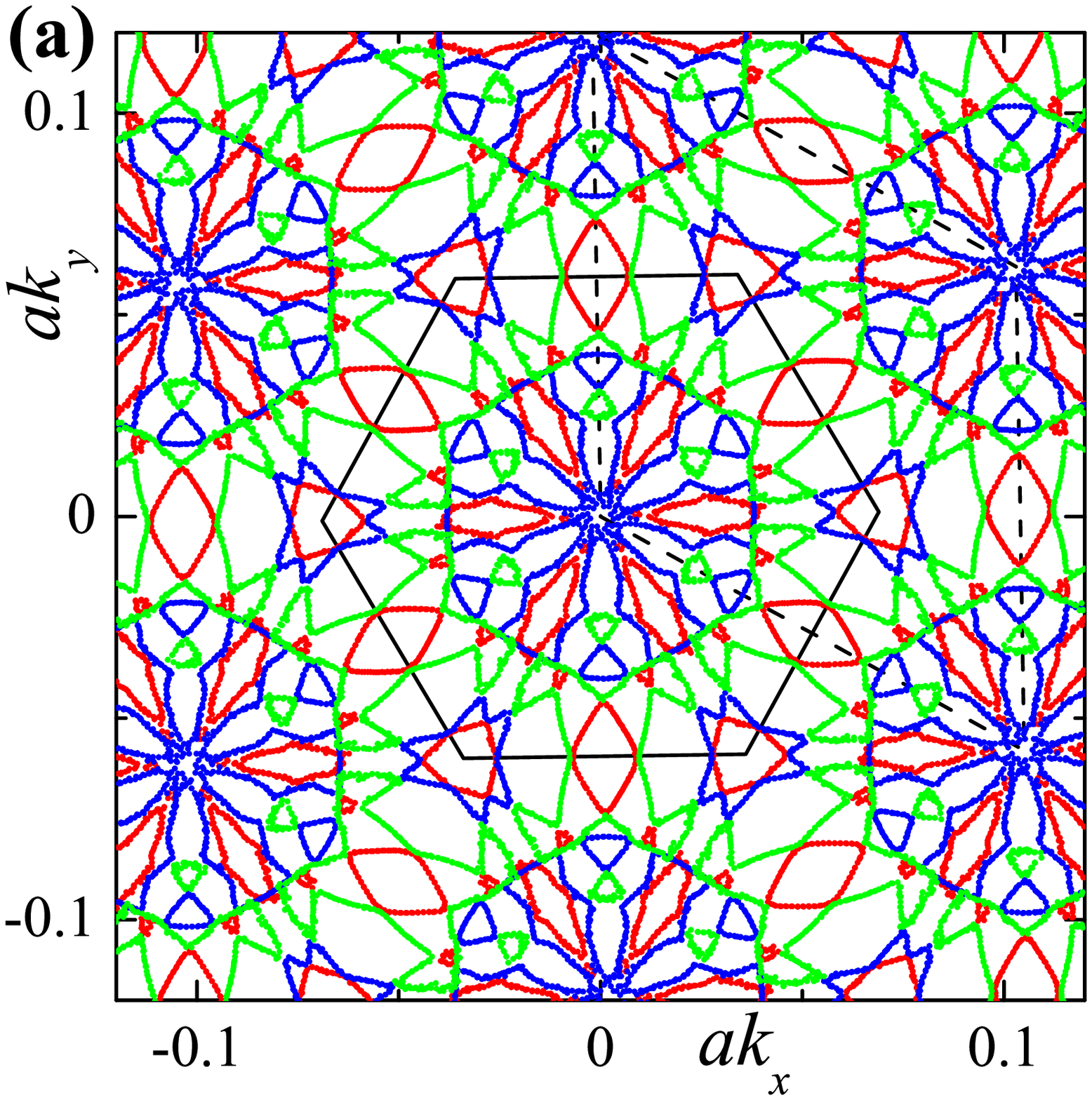}\hspace{8mm}
\includegraphics[width=0.46\textwidth]{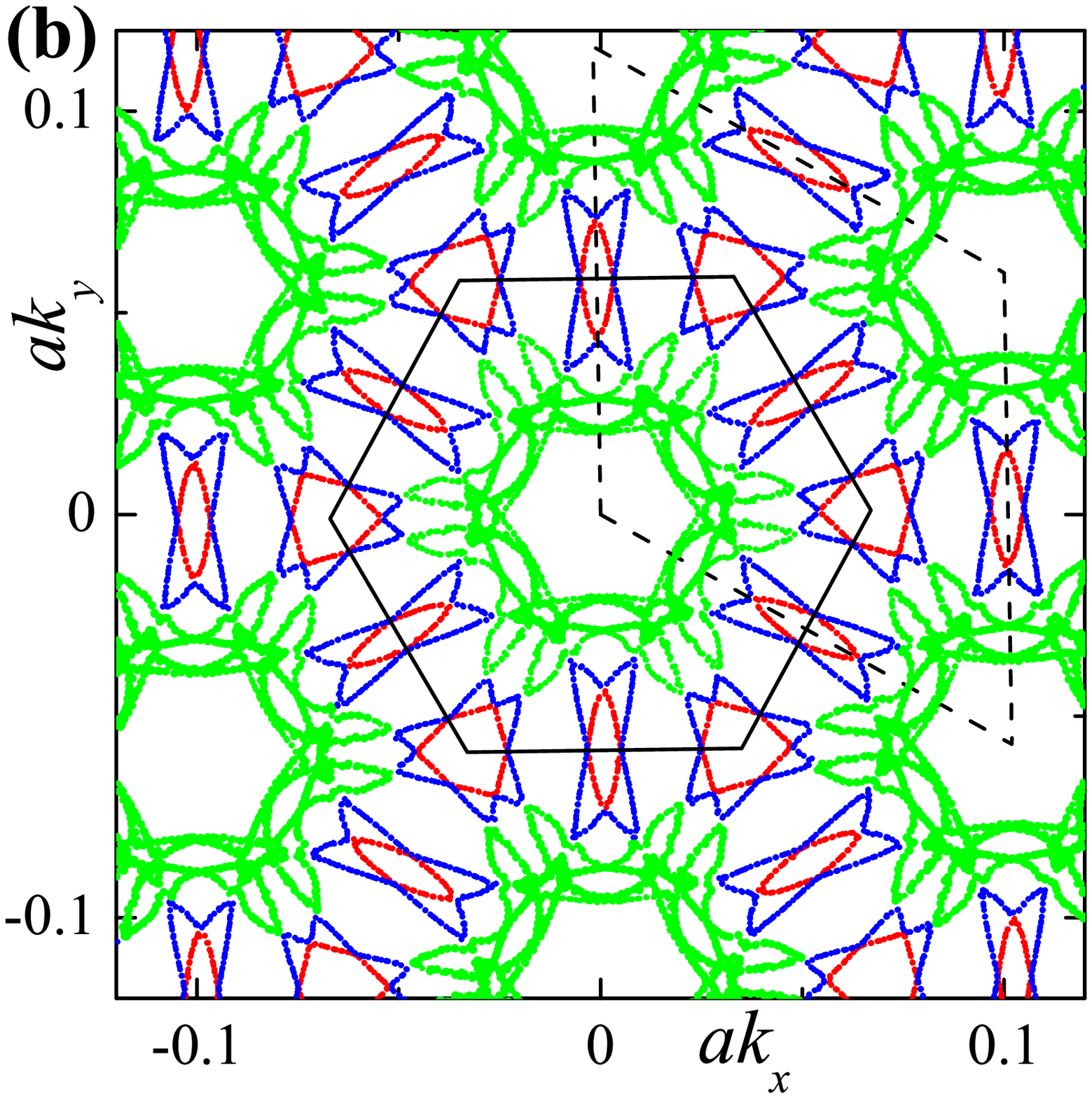}\\
\includegraphics[width=0.46\textwidth]{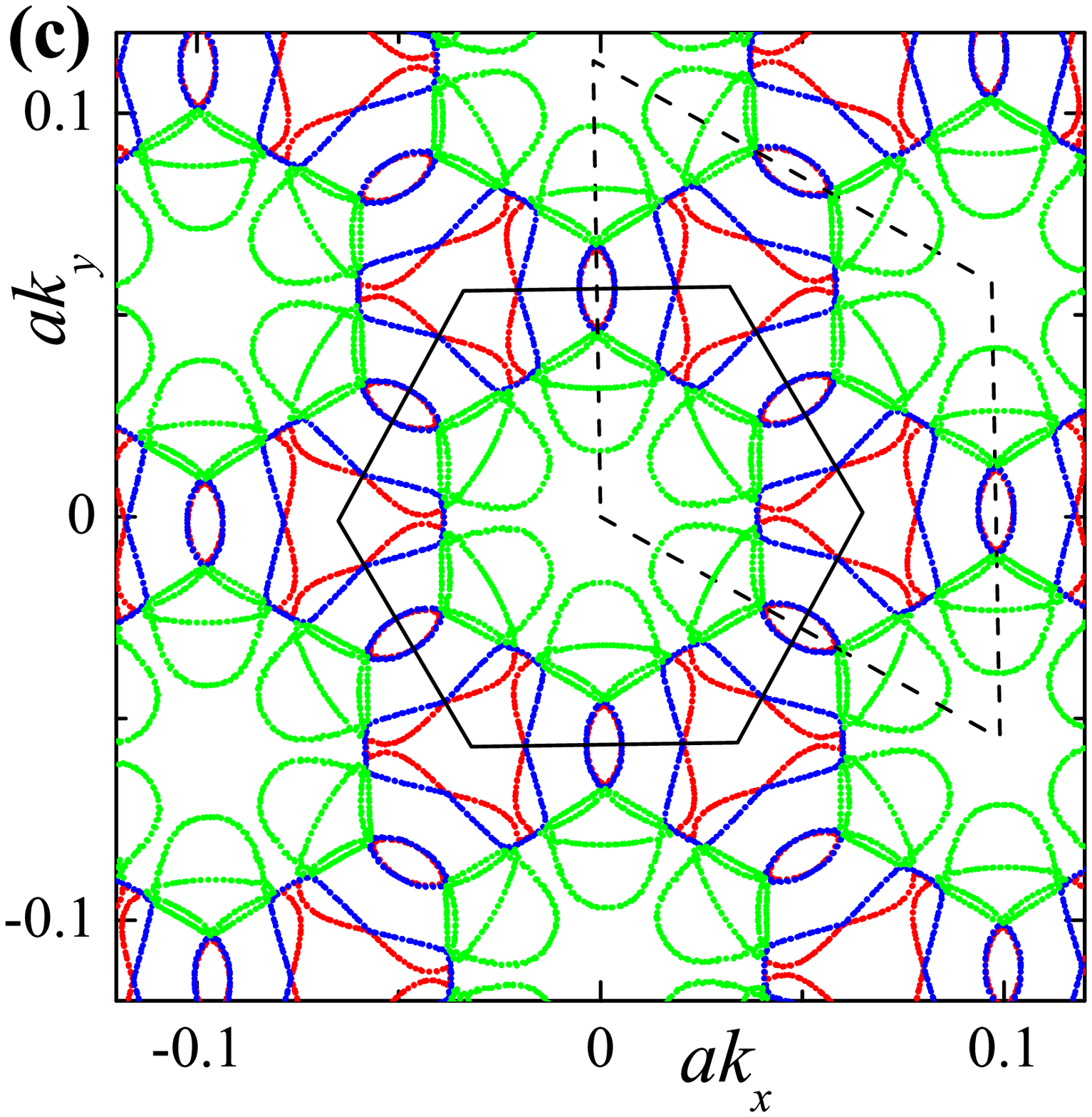}\hspace{8mm}
\includegraphics[width=0.46\textwidth]{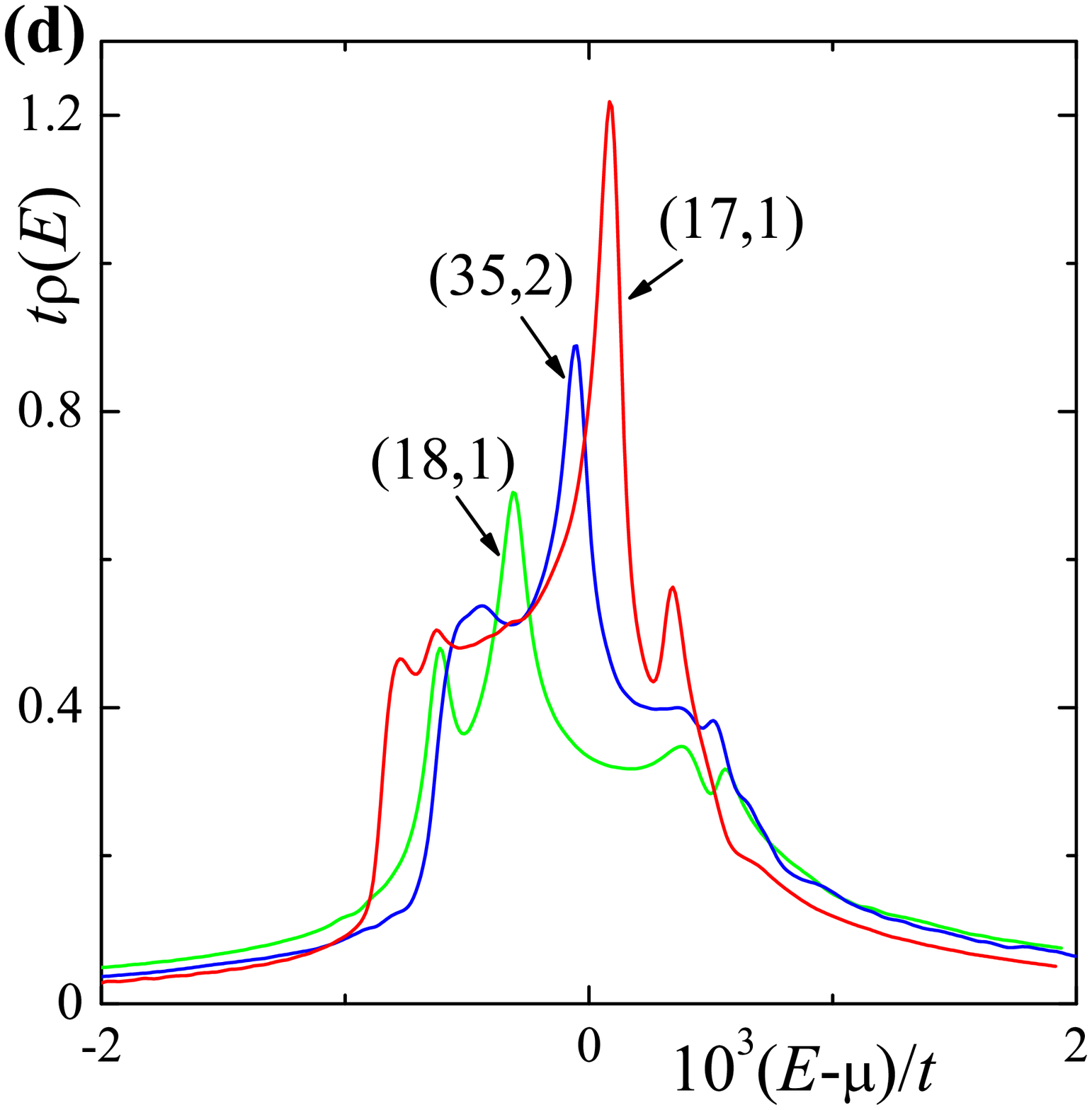}
\caption{(Color online) Fermi surfaces of the superstructures $(17,1)$
[$\theta\cong1.89^{\circ}$, panel (a)],
$(35,2)$
[$\theta\cong1.84^{\circ}$, panel (b)],
and $(18,1)$
[$\theta\cong1.79^{\circ}$, panel (c)]
calculated at half-filling. The Fermi surfaces for the structures
$(17,1)$
and
$(18,1)$
are calculated by band folding of the original Fermi surfaces. Different colors
correspond to different bands intersecting the Fermi level $\mu$ at
half-filling. Panels~(a) and~(c) show the same Fermi surfaces as
panels~(a) and~(b) of
Fig.~\ref{FigFSr1},
but in the folded Brillouin zone.
(d) The low-energy density of states calculated for three
superstructures with similar twist angles
$\theta<\theta_c$.
The density of states is calculated at finite temperature
$T/t=10^{-5}$
by numerical integration over the momentum, see
Eq.~\eqref{rhoT}.
Note that in units of $t$ the energy window where the weight is
enhanced is very narrow.
}\label{FigFS}
\end{figure*}

The calculations for samples of different sizes
($\tilde{L}/a=61,\,75,\,151$)
show that the characteristic width of the peak,
$\overline{\Delta\theta}$,
satisfies the relationship
$$
\overline{\Delta\theta}\tilde{L}/a\sim1\,.
$$
In other words, the gap disappears when the maximum deviations in the
positions of the atoms in the sample
$\tilde L \delta \theta$
is about a lattice constant. The larger the value of
$\tilde{L}$,
the smaller fluctuations in the twist angle destroy the band gap.

We expect that similar results are also valid for an infinite, but non-ideal
sample with a finite mean-free path of electrons. If this so, we would have
a paradoxical situation: the less defects the sample has, the more difficult would be to
experimentally observe the gap in the spectrum due to the fluctuations of
the twist angle.  This issue, as well as the study of the effects of other
possible fluctuations in the tBLG crystal lattice are beyond the scope of this work.

\section{Small twist angle $\theta<\theta_c$}
\label{sect::small_angle}


The Fermi velocity $v_F$ calculated according to Eq.~\eqref{Efit} decreases
when $\theta$ decreases (see
Fig.~\ref{FigDeltaV}),
in good agreement with previous
theoretical~\cite{Pankratov1,Pankratov2,Pankratov3,NanoLett,Morell,dSPRL,dSPRB,PNAS}, and experimental~\cite{STM_LL,Raman1} studies.
For angles close to $\theta_c=1.89^{\circ}$, four low-energy bands
become almost flat in the whole Brillouin zone, with the exception of the
small region near the point ${\bm\Gamma}$ [see Fig.~\ref{FigSpec}(a--c)].
In the region of twist angles
$\theta_c\cong1.89^{\circ}<\theta\lesssim4.4^{\circ}$,
the band splitting is too small to be experimentally observable even for
$r=1$ superstructures.
Since the band splitting is negligible, the electronic structure changes
continuously

For smaller angles,
$\theta<\theta_c$
($m_0\geqslant17$),
the cone-like shape of the low-energy bands completely disappears even near
the Dirac points [see
Fig.~\ref{FigSpec}(f)],
the gap becomes zero, and the system acquires a Fermi surface and non-zero
density of states at the Fermi level. As $\theta$ decreases further, the
Fermi surface changes, and for some values of $\theta$ reduces to several
Fermi points. The density of states at the Fermi level oscillates with
$\theta$.

Figure~\ref{FigFSr1}
shows the Fermi surfaces calculated at half-filling for the superstructures
$(17,1)$ corresponding to
$\theta_c\cong1.89^{\circ}$
and similar to the structure
$(18,1)$
($\theta\cong1.79^{\circ}$). It is clearly
seen from these figures that the Fermi surfaces are similar to each other,
and the total size (length) of the Fermi surface sheets for the structure
$(18,1)$ is smaller than that for the $(17,1)$.
The band flatness, the non-zero density of states for small $\theta$, as
well as the existence of the `magic' angles where the density of states
vanishes is consistent with many previous studies using both
low-energy~\cite{dSPRB,PNAS,NonAbelianGaugePot}
and tight-binding
calculations~\cite{Pankratov1,NanoLett,TramblyTB_Loc,Morell}.

Our calculations show that no gap exists between the low-energy bands
and the lower or upper bands. Thus, the system remains metallic under doping
when the chemical potential shifts from its values at half-filling. On the other hand, the bands $\bar{E}_{\mathbf{k}}^{(\nu)}$
are quite flat if $\theta<\theta_c$; the Fermi velocities are about $10^{-3}$ times smaller than
for a single graphene layer. Consequently, the disorder or the
electron-electron interaction may qualitatively change the metallic band
structure giving rise to localization or opening of a gap due to ordering.


The superstructures with
$r\ne 1$
are also metallic if $\theta$ is smaller than
$\theta_c$.
To compare the Fermi surfaces, we perform band folding for
$r=1$
superstructures as described in
subsection~\ref{SectionRneq1}.
Figure~\ref{FigFS}
shows the Fermi surfaces for the superstructures $(17,1)$
[$\theta\cong1.89^{\circ}$, panel (a)] and $(18,1)$
[$\theta\cong1.79^{\circ}$, panel (c)] calculated in the folded (reduced
$2$ times in size) Brillouin zones, as well as the Fermi surface for the
intermediate $(35,2)$ [$\theta\cong1.84^{\circ}$, panel (b)]
superstructure. Since the twist angles are quite small, the
Brillouin zones considered almost coincide with each other. All Fermi surfaces are
calculated at half-filling, and the position of the chemical potential
$\mu$ in each case is found in the way described in Section~\ref{TBM}. We
see, that the Fermi surface of the intermediate $r=2$ superstructure, being
different in some details, contains, however, all basic elements presented
in the Fermi surface (in folded Brillouin zones) of both proximate
$r=1$ superstructures.

For a more quantitative analysis, we calculate the low-energy density of
states of $r=1$ and $r\neq1$ superstructures with similar twist angles. The
densities of states near the Fermi level for the $(17,1)$, $(35,2)$, and
$(18,1)$ superstructures are shown in Fig.~\ref{FigFS}(d). Each density
of states has a sharp peak and a shoulder, which has addition smaller
peaks. The height and the position of the central peak with respect to the
Fermi level, as well as the height and the position of the shoulder
correlate with the change of the twist angle. For the superstructure $(m_r,r)$
with $\theta<\theta_c$, the number of low-energy bands which contribute to
the peaks and shoulder in the density of states at low energies is equal to
$n_0=4r^2/g$, where $g=1$ if $r\neq3n$ or $g=3$ otherwise. With the
normalization of the DOS used in Eq.~\eqref{rhoT}, the integral of $\rho_T(E)$
over low energies [the area under the curves shown in
Fig.~\ref{FigFS}(d)] with a high accuracy is equal to
$S=4r^2/[gN(m_r,r)]$. Using Eqs.~\eqref{theta} and~\eqref{Nsc}, one can
easily show that this integral depends only on the twist angle and is equal
to \begin{equation}
S=\sin^2\!\frac{\theta}{2}\,.
\end{equation}
The spectral weight shifts toward higher energies as the twist angle
decreases.

Thus, our analysis of the DOS and the Fermi surface indicates that
neglecting some delicate details, the electronic properties of the tBLG
change continuously with the twist angle when
$\theta<\theta_c$.
However, further analysis is required to address the issues of the evolution of the
Fermi surface at very small angles (e.g., the existence of the `magic' angles where
the Fermi surface vanishes, etc.).

\section{Conclusions}
\label{sect::conclusions}
To conclude, we have studied a tight binding model for twisted bilayer
graphene in a wide range of twist angles. In the model Hamiltonian we
take into account the effect of the environment-dependent hopping, which
correctly reproduces the Slonczewski-Weiss-McClure scheme for inter-layer
hopping amplitudes in bilayer graphene. We demonstrate that at twist
angles
$\theta>\theta_c\cong1.89^{\circ}$
the tBLG can have a band gap, which can be as large as
$80$\,meV.
The gap is maximum for twist angles corresponding to superstructures with
the superlattice period coinciding with their Moir\'{e} period. This gap,
however, is very sensitive to small deviations of the twist angle from
these original values. This sensitivity of the gap disappears for
finite-size samples. If $\theta$ is below a critical angle
$\theta_c$, tBLG has a Fermi surface, and the DOS has a peak at the Fermi level.
Moreover, the DOS changes continuously with the twist angle.

\section*{Acknowledgments.}

The authors are grateful to A.N.~Sergeev-Cherenkov for computing assistance. This work was supported in part by the RFBR (Grants Nos.~14-02-00276, 14-02-00058, 12-02-00339), the RIKEN iTHES Project, MURI Center for Dynamic Magneto-Optics, and a Grant-in-Aid for Scientific Research (S).


\vspace*{-0.1in}
%


\begin{thebibliography}{25}
\bibitem{note1}
We take $R_{ij}=r_{ij}$ in Eq.~(2) of Ref.~\onlinecite{Tang} since,
with a high accuracy, the effective coordination numbers $g_i$ in Eq.~(4) of
Ref.~\onlinecite{Tang}  turn out to be the same for all sites.
As a result, we obtain $R_{ij}=Cr_{ij}$, and the constant $C$ is
absorbed in the coefficients $\alpha_1$ and $\alpha_4$.
\end{thebibliography}

%

\end{document}